\documentclass[11pt]{article}
\input epsf.sty
\newcommand{\ra}{\rangle}
\newcommand{\la}{\langle}

\newcommand{\ta}{{\cal T}}

\newcommand{\HH}{{\cal H}}

\newcommand{\LL}{{ L'^ {g}}}

\newcommand{\TT}{{\cal T}}

\newcommand{\GG}{{\cal G}}

\newcommand{\be}{\begin{equation}}
\newcommand{\ee}{\end{equation}}
\newcommand{\ben}{\begin{eqnarray}\displaystyle}
\newcommand{\een}{\end{eqnarray}}
\newcommand{\refb}[1]{(\ref{#1})}

\newcommand{\sectiono}[1]{\section{#1}\setcounter{equation}{0}}

\begin{document}

{}~
\hfill\vbox{\hbox{hep-th/0211012}
\hbox{PUPT-2054}
}\break

\vspace{2cm}

\centerline{\Large \bf
Experimental String Field Theory
}

\vspace*{8.0ex}

\centerline{\large \rm Davide Gaiotto  and Leonardo Rastelli }

\vspace*{6.0ex}

\centerline{\large \it  Physics Department, Princeton University, Princeton, NJ 08544}

\vspace*{2ex}

\centerline{E-mail: {  \tt dgaiotto,  lrastell@princeton.edu } }

\vspace*{1.8ex}

\vspace*{10.0ex}

\centerline{\bf Abstract}

\smallskip

\smallskip

\smallskip

We develop efficient algorithms
for level-truncation computations 
in open bosonic string field theory.
We  determine the classical
action in the universal subspace
to level (18,54) and apply
this knowledge  to numerical
evaluations of the tachyon condensate
string field. We obtain two main sets of results. 
First, we directly compute the 
solutions up to level $L=18$
by extremizing the level-truncated action.
Second, we obtain  predictions
for the solutions for $L > 18$ from an extrapolation to higher levels
of the functional form of the tachyon effective action.
We find that the energy  of the stable vacuum overshoots  -1 (in units of the brane
tension) at $L=14$, reaches a minimum $E_{min}  =  -1.00063$ 
at $L \sim 28$  and approaches with spectacular accuracy
the predicted answer of   -1  as $L \to \infty$.  
Our data are entirely consistent 
with the recent  perturbative analysis of Taylor
and strongly  support the idea that level-truncation is a convergent approximation
scheme.
We also  check systematically that our
numerical solution, which obeys the Siegel gauge
condition, actually satisfies  the full gauge-invariant equations
of motion.
Finally we investigate the presence of analytic patterns
in the coefficients of the tachyon string field,
which we are able to reliably estimate  in the $L \to \infty$ limit.

\vspace{4cm}

\vfill \eject

\baselineskip=16pt

\tableofcontents

\newpage

\sectiono{Introduction and Summary} \label{s1}

The realization that
D-branes are solitons of the open string tachyon \cite{conj, 9911116}
has triggered a revival of interest in open string field theory.
Much work has focused on the search for  classical solutions
of cubic bosonic open string field theory  \cite{OSFT} 
(OSFT). Despite important technical
progress in the understanding of the open
string star product - notably  the discovery of  star algebra projectors \cite{0006240, projectors}
and of new connections with non-commutative
field theory \cite{bars, spectroscopy, 0202087, other*} -
analytic classical solutions  of OSFT are still missing\footnote{Exact results have been obtained in vacuum string
field theory \cite{vsft1, vsft2} (VSFT), see {\it e.g.} \cite{projectors, okawa,
HK, exactvsft, strings2002}.
VSFT is the version of open string field theory
which appears to describe the stable tachyon vacuum.
In VSFT the BRST operator is replaced by a purely ghost
insertion at the string midpoint and
the equations of motion take the exactly solvable
form of projector equations 
(with an auxiliary `twisted' ghost system \cite{vsft2}).
It would nevertheless be extremely desirable to solve 
analytically the original
OSFT equations. See {\it e.g.} \cite{attempts} 
for some formal attempts.}.

\smallskip

Fortunately, the  OSFT equations of motion can
be solved {\it numerically}
in the `level-truncation' scheme
invented by Kostelecky and Samuel \cite{KS}.
 The  open string field is restricted
to modes with an $L_0$ eigenvalue smaller than a prescribed maximum
`level' $L$. For any finite $L$, the truncated OSFT action
contains a finite number of fields and numerical computations
are possible. Remarkably, numerical results \cite{KS}-\cite{patterns}
for various classical solutions 
appear to converge rapidly to the expected answers as
the level $L$ is increased.  
Much of our present intuition
about the classical dynamics of OSFT comes from
the level truncation scheme, and in fact even in
vacuum string field theory \cite{vsft1, vsft2}
  several exact results were first guessed based on numerical
data.

\smallskip

This motivated  us to develop efficient algorithms for  level-truncation calculations.
 Our main technical
innovations are the systematic use of conservation laws \cite{0006240}
to compute the cubic vertices, and the implementation of our algorithms 
on a C{++} code. 
In this paper
we apply these methods to the evaluation of the
classical action in the {\it universal} subspace \cite{9911116, 0006240},
which is the space of string fields  
generated by ghost oscillators and matter Virasoro
generators acting on the vacuum.
Using conservation laws we 
 determine the classical action directly in the universal
basis in a recursive way, with an algorithm
whose complexity
is linear  in the number of vertices
(cubic  in the number of fields). Some
details about
the numerical algorithms can be found in 
appendix A of the paper.
The gain in efficiency of our methods 
is of several orders of magnitude, and we
are able to obtain the $10^{10}$ 
universal cubic vertices at level (18,54).

\smallskip

The universal subspace
has special physical significance because
it contains the tachyon condensate string field,
the solution of OSFT corresponding to the stable
vacuum of the open string tachyon. 
Its (negative) energy per unit volume
must exactly cancel the D-brane tension.
Sen and Zwiebach's computation \cite{SZ} of the tachyon condensate up to
level (4,8)  gave the first evidence that OSFT reproduces the correct
D-brane physics.  Moeller and Taylor \cite{MT}
pushed
the computation to level (10,20) finding that
99.91\% of the D-brane tension is cancelled in the tachyon vacuum.
 Given such a remarkable agreement, it may appear 
quite pointless to extend their
results to higher level. Not so. 
Up to level 10,
the individual coefficients of the string field
appear to converge much less rapidly
than the value of the action. 
A more precise determination of the coefficients
is likely to provide clues for an exact solution. Indeed
various surprising patterns obeyed by OSFT solutions
were `experimentally' observed in \cite{patterns}, but
more accurate results are needed to decide
which of these patterns are likely to be exact and which
ones only approximate. Higher level computations
can also be expected to shed light on the nature
of the level truncation procedure itself,
which still lacks a sound theoretical justification.

\smallskip

Our first main set of results (described in section 3)
is the computation  of the Siegel gauge tachyon condensate in 
level-truncation up to $L=18$. 
The procedure is the standard one: at any given level $L$, 
there are $N_L$ scalar fields that obey
the Siegel condition, and we determine
their vev's by solving the
 $N_L$ equations of motion implied by the gauge-fixed
action. There is a potential subtlety here:
the full  equations of motion before gauge-fixing 
impose a bigger number of constraints \cite{HS} 
(the extra conditions simply enforce extremality of the action along  gauge
orbits). Consistency demands that the full set of equations of motion is satisfied as $L \to \infty$,
and we systematically check that this indeed happens,
with remarkable accuracy. As another consistency check,
we verify that the tachyon condensate obeys
the quadratic relations analytically derived
by Schnabl \cite{schnabl}, 
and we again find excellent agreement.

\begin{table}
\begin{tabular}{|l|l|l|}
\hline
$L$ & $E_{(L, 3L)}$ & $E_{(L,  2L)}$  \\
\hline
\hline
2  & -0.9593766 &  -0.9485534   \\
\hline
4 & -0.9878218 &  -0.9864034     \\
\hline
6 & -0.9951771 &    -0.9947727   \\
\hline
8   &  -0.9979302&-0.9977795  \\
\hline
10 &   -0.9991825 &  -0.9991161   \\ 
\hline
12 & -0.9998223  &  -0.9997907 \\
\hline
14 &-1.0001737  &-1.0001580 \\
\hline
16 & -1.0003754   & -1.0003678  \\
\hline
18 &  -1.0004937 &  -1.00049   \\
\hline
\end{tabular}
\caption{\label{energies}
Values of the vacuum energy  in level-truncation,
in the $(L,3L)$ and $(L,2L)$ approximation schemes. }

\end{table}
\smallskip

The values of the vacuum energy
as a function of $L$ are shown in Table \ref{energies}. Unexpectedly,
at $L=14$ the energy overshoots the predicted answer
of -1 and appears to further decrease 
at higher levels.  As a first reaction,
one may wonder whether the level-truncation procedure
is breaking down for $L > 10$, as
could happen if the approximation was only asymptotic.
In this pessimistic scenario, for any OSFT observable 
there would be a maximum level that gives the estimate
closest to the `exact' value, and beyond
this optimal level the procedure would stop converging.
 However, the data favor a smooth
behavior as $L$ increases,
since the differences between consecutive
approximations are getting smaller.

\smallskip

The results in Table 1 may simply indicate that
the approach of the energy to -1 as $L \to \infty$ is non-monotonic,
contrary to previous naive expectations.
 Indeed, Taylor has recently presented convincing evidence \cite{wati}
for
this benign interpretation of our results.\footnote{The data in Table 1 
 were first announced  at the Strings 2002 conference, Cambridge,  July 2002 \cite{strings2002} .} 
He applies a clever extrapolation technique to 
level-truncation data for $L \leq 10$
 to estimate the vacuum energies even for $L > 10$. 
This procedure
 reproduces quite accurately  our exact values in Table 1 and  further
predicts that the vacuum energy reaches a minimum for
 $L \sim 28$, but then turns back to approach asymptotically
-1 for $L \to \infty$.

\smallskip

\smallskip
 \begin{figure}[!ht]
\leavevmode
\begin{center}
\epsfysize=9cm
\epsfbox{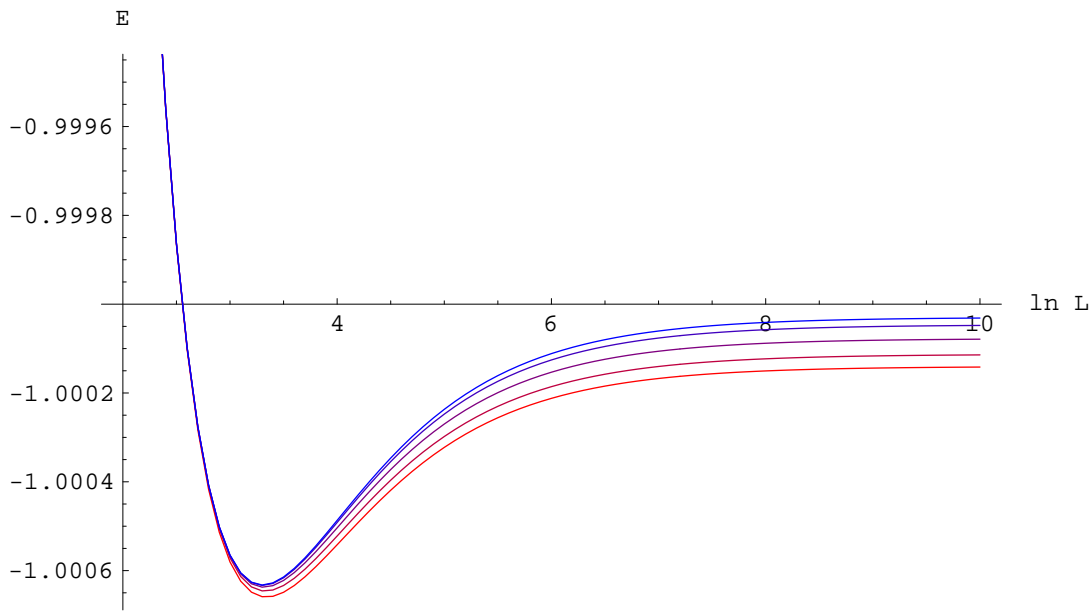}
\end{center}
\caption{Curves of the vacuum energy  as a function of level,
as predicted by our extrapolation scheme  for
various values of $M$ (maximum level of the data
used in the extrapolation). The figure shows the curves
 $E^{(M)}(L)$  on a logarithmic plot,
for $M$ between 8 (lowermost curve) and 16 (uppermost curve). Data in
the $(L,3L)$ scheme.}
\label{logactions} 
\end{figure}

In sections 4 we introduce our second main set of results.
We devise
an extrapolation technique in the same spirit of Taylor's analysis.
We consider the effective tachyon potential  $V_L(T)$ around the unstable vacuum, 
obtained by classically integrating out  all the  higher scalars up to  level $L$.
$V_L(T)$ is computed `non-perturbatively' by fixing
the value of $T$ and solving numerically the equations
of motion for  the other scalars\footnote{In this we differ
from Taylor \cite{wati}, who uses instead a series expansion
of the potential in powers of $T$.
}. 
 We are able to
obtain $V_L(T)$ up to $L=16$.
 Clearly for each $L$, the minimum of $V_L(T)$ is just the vacuum
energy $E_L$ at level $L$. However the functional dependence
on $T$ contains more information than just
the extremal value $E_L$.
The idea is to perform an extrapolation in $L$ of
the whole {\it functions} in $T$.  
In practice, we consider a finite interval of values
of $T$ around the non-perturbative minimum. 
For  a fixed $T$  in this interval 
we interpolate our data for $V_L(T)$  with a  polynomial in $1/L$, 
and then extrapolate this polynomial  to higher levels. 
To check the stability
of this approximation scheme, we vary the maximum
level $M$ of the set of data 
used as input for the extrapolation:
for each $M \leq 16$, we apply
the extrapolation to
the functions $\{  V_L(T)  \, |\, L \leq M \}$. This gives
estimates
 $T_L^{(M)}$ for the tachyon vev and
for the corresponding vacuum energy
$E^{(M)}_L$, for any $L > M$.

\begin{table}
\begin{tabular}{|l|l|l|l|}
\hline
     &$L_{min}$&$E_{min}$&$E_{\infty}$ \\
\hline
$M=6$&41.1&-1.001171&-1.000949 \\
\hline
$M=8$&28.2&-1.000660&-1.000140 \\
\hline
$M=10$&27.8&-1.000646&-1.000113 \\
\hline
$M=12$&27.5&-1.000637&-1.000077 \\
\hline
$M=14$&27.3&-1.000633&-1.000046 \\
\hline
$M=16$&27.3&-1.000632&-1.000030 \\
\hline
\end{tabular}
\caption{\label{Einf}
Parameters of the curves $E^{(M)}(L)$ (in the $(L,3L)$ scheme).
The energy reaches its minimum $E_{min}$ for $L = L_{min}$,
and tends asymptotically to $E_\infty$ as $L \to \infty$.}
\end{table}

\smallskip

The predicted power of the method is quite
impressive. For example, with $M=10$, that
is using only level-truncation results up to level 10,
the estimate $T^{(10)}_{18}$ reproduces with an
accuracy of $10^{-5}$  the exact tachyon vev $T_{18}$,
obtained  by straightforward level-truncation  at $L=18$.
This is remarkable, since the former computation
is  over  a  thousand times faster than the latter.

\smallskip

Figure \ref{logactions} and Table \ref{Einf} summarize
the extrapolations of the energy as a function of level, 
for various values of $M$. The data completely confirm
(with  enhanced precision) the conclusions of  Taylor \cite{wati}.
The behavior of the energy as a function of level
is non-monotonic, but eventually the asymptotic
limit of -1 is reached with spectacular
accuracy.

\smallskip

 These results
greatly reassure us of the validity of the level-truncation scheme.
Observables in OSFT  
have a smooth limit 
as $L \to \infty$, which (in the absence of an alternative definition)
should be identified with their  `exact'  value. In all cases where
 an independent prediction for the observable is available
(as for  the vacuum energy, or for Schnabl's quadratic identities),
the $L \to \infty$ extrapolation gives the correct answer.

\smallskip

A practical lesson of this analysis is that
polynomial interpolations in $1/L$ have
 great predictive power,
at least for the $(L,3L)$ approximation 
scheme\footnote{
In the $(L,3L)$ scheme, the string field is truncated up to level
$L$ and all of its mode are kept in the OSFT action. By contrast,
in the $(L,2L)$ scheme one keeps only the cubic terms in the action
whose total level is $\leq 2L$. As discussed in section \refb{L2L},
$(L,3L)$ results display a much smoother dependence on $L$. }.  
This observation makes the level-truncation scheme much more
efficient, as reliable estimates can be extracted from
(relatively) painless numerical work.

\smallskip

In section 5 we describe the results for the individual
coefficients of the tachyon string field extrapolated to $L = \infty$.
The main conclusions are:

\begin{itemize}

\item
The asymptotic value of the tachyon vev is $T_{\infty} \simeq 0.5405$,
 ruling out the conjecture \cite{HS}  for an exact  value $T_{HS} = \sqrt{3}/\pi \cong
0.5513$.  

\item
The conjectured universality \cite{patterns}
of certain ghost coefficients
seems to be somewhat problematic in view of our data.
While most coefficients
come strikingly close to the conjectured values,
the coefficient of $c_{-1} | 0 \ra$ deviates by $2 \%$ 
from the expected answer.
Our predictions for the $L=\infty$ string field are likely
to have a smaller error. 

\item
The `quasi-pattern' observed in \cite{patterns} of an approximate
factorization of $L_0 \TT_{Siegel}$, where $\TT_{Siegel}$ is the Siegel-gauge
tachyon condensate, gets worse in our $L =\infty$ extrapolations.
So this is definitely not an exact property.

\item The  correspondence between
OSFT equations of motion in Siegel gauge 
and a certain equation for ghost number zero
string fields \cite{patterns} is still rather well obeyed,
but does not  improve in the $L = \infty$ extrapolations.
 This is likely an approximate pattern,
as already suspected in \cite{patterns}.

\end{itemize}

We hope that our accurate data
will stimulate new imaginative approaches to the problem
of finding an exact solution.   
In the near future it will  be possible to
extract from our results new  information about the kinetic term around the tachyon vacuum.
It will also be straightforward to extend the methods of this paper to the computation of
more general classical solutions of OSFT, which  should provide more analytic clues.

\smallskip

A more complete set of numerical data 
that it is practical to reproduce here (coefficients at higher levels, 
more significant figures etc.) will be be made available on-line \cite{webpage}.

\smallskip
To make this paper self-contained,
we begin in the next section with a review 
of some basics.

\sectiono{OSFT and the Universal Tachyon Condensate}

In this section we describe
the basic setup
for classical equations of motion in OSFT,
with an emphasis on the symmetries obeyed by 
the tachyon condensate string field in Siegel gauge.
While none of the results 
presented in this section are really new,  we 
take the opportunity to spell out some
facts - in particular our summary equation \refb{siegelsingl} -
that are not widely known.

\subsection{The Tachyon Condensate}

The action of OSFT takes the well-known (deceptively) simple form
\cite{OSFT}
\be \label{Waction}
S = -\frac{1}{g^2} \left( \frac{1}{2} \la \Psi, Q_B \Psi \ra +
\frac{1}{3} \,\la \Psi, \Psi * \Psi \ra \right)\,.
\ee
This action  describes the worldvolume dynamics of
a D-brane specified by some Boundary CFT. The
string field $\Psi$ belongs to the full matter+ghost state-space of this 
BCFT. In classical OSFT, $\Psi$ has ghost number one\footnote{
Our conventions and notations are the same as \cite{0006240}. 
In particular
we define the SL(2,R) vacuum $|0 \rangle$ to
have ghost number zero, and the ghost and antighost fields $c(z)$ and $b(z)$
to have ghost number one and minus one, respectively.}.
According to Sen's conjecture \cite{conj},  the classical
OSFT eom's  
\be \label{OSFTeq}
Q_B \Psi + \Psi * \Psi = 0 \, 
\ee
must admit a Poincar\'e  invariant solution $\Psi \equiv \ta$ 
corresponding
to the condensation of the open string tachyon to the vacuum with 
no D-branes. The tachyon potential ${\cal{V}}(\Psi)$ is given 
by \cite{9911116}
\be \label{fPsi}
{{\cal V}(\Psi)\over 2 \pi^2 M} \equiv \frac{1}{2 \pi^2} f(\Psi) =
\,{1\over 2}
\langle \Psi, Q_B  \Psi \rangle + {1\over 3}\, \langle\, \Psi , \Psi * 
\Psi \rangle \,,
\ee
where $M$ is the brane mass. The normalized potential
$f(\Psi)$ is expected to equal minus one at the tachyon vacuum,
\be
f(\ta) = -1 \,.
\ee

\subsubsection{Universality}
A basic property of the tachyon condensate string field $\ta$ is
{\it universality} \cite{9911116},
\be 
 \ta \in {\cal H}_{univ}^{(1)} \, ,
\ee
where 
\ben \label{universal}
 \HH_{univ} \equiv {\rm Span} \{ L^m_{-j_1} \dots L^m_{-j_p}\,
b_{-k_1} \dots b_{-k_q} \, c_{-l_1} \dots c_{-l_r} | 0 \ra\, , 
 \; j_i \geq 2, k_i \geq 2, l_i \geq -1 \} \, 
\een  
with $L^m_k$ denoting the matter Virasoro generators.
The universal space is further decomposed
into a direct sum of spaces with definite ghost number
\be
\HH_{univ} = \oplus_{n\in Z} \HH_{univ}^{(n)}\,.
\ee
The restriction of the classical action to $\HH_{univ}$ can
be  evaluated using purely combinatorial algorithms
that only involve the ghosts and the matter
Virasoro algebra with $c=26$ \cite{9911116, 0006240}. It follows
that the form of $\ta$ does not depend
on any of the details of the BCFT that defines the D-brane background
before condensation.

\subsubsection{Twist}

An obvious symmetry of the tachyon condensate is twist
symmetry. 
The OSFT equations of motion admit a consistent truncation
to twist even string fields \cite{GZ}, and  indeed the tachyon condensate
solution turns out to be twist even. 
In $\HH_{univ}^{(1)}$, twist is
defined simply as $(-1)^{L_0 +1}$, so  $\TT$ contains 
only states with {\it even} level $L \equiv L_0 +1$.

 \subsubsection{Siegel gauge and SU(1,1)}

The Siegel gauge condition 
$ b_0 \Psi = 0 $
is particularly natural in level truncation
since it is easily imposed level by level
by simply   omitting all Fock states containing the ghost zero mode $c_0$.

\smallskip

The Siegel  gauge-fixed equations of motion 
\be \label{Siegelequ}
L_0 \Psi + b_0 (\Psi * \Psi) = 0
\ee
admit a consistent truncation to the subspace of string
fields which are singlets of SU(1,1) \cite{0010190}. The SU(1,1)
symmetry in question is generated by
\be
\GG = \sum_{n=1}^\infty \left( c_{-n} b_n - b_{-n} c_n \right) \quad
X =   -      \sum_{n=1}^\infty \left( n \, c_{-n} c_n \right)
\quad Y =    \sum_{n=1}^\infty \left( 
\frac{1}{n}  b_{-n} b_n  \right)\,
\ee
and the singlet subspace is defined as
\be   \label{singletdef}
\Psi  \in \HH_{singl} \quad {\rm iff} \; b_0 \Psi = \GG \Psi  = 
X     \Psi = Y \Psi = 0 \,.
\ee
Notice  that   acting on Siegel states
$\GG$   is just ghost
number  shifted by one unit,
so  all states in $\HH_{singl}$ have ghost number one. 
To  show consistent truncation of equations
\refb{Siegelequ} to the singlet subspace, 
we need to prove
that if $\Psi \in \HH_{singl}$, then  $b_0 (\Psi * \Psi) \in \HH_{singl}$,
so that all components
of $\Psi$ outside $\HH_{singl}$ can be consistently set to zero.
A simple argument is as follows. The generator $X$ is a derivation
of the $*$-algebra\footnote{It is enough to notice that $-2 X = 
\{ Q_B, c_0 \}$, see \refb{Q} below.
Both $Q_B$ and $c_0$ are derivations \cite{0006240}, and
(anti)commutators of (graded) derivations are derivations.
On the other hand, the generator $Y$ is {\it not} a derivation.}, 
and commutes with $b_0$. Hence if $X \Psi =0$,
$X \, b_0 (\Psi * \Psi) = 0$. Clearly $\GG$ is also zero on 
$b_0 (\Psi * \Psi)$, since ghost number adds
under $*$-product. By the structure of the 
finite-dimensional\footnote{Since
SU(1,1) commutes with $L_0$, we can run the argument in the
subspaces of $\HH_{univ}$ with given $L_0$, which are finite-dimensional.}
representations of SU(1,1), 
a vector
with zero $\GG$ and $X$ eigenvalues must also
have zero $Y$ eigenvalue, that is,  $b_0 (\Psi * \Psi) \in \HH_{singl}$,
as desired.

\smallskip

The SU(1,1) singlet subspace has a simple characterization in terms
of the  Virasoro generators of the `twisted' ghost conformal
field theory of central charge $-2$ \cite{vsft2} 
\footnote{ 
A proof of  the equivalence of definitions \refb{singletdef} and \refb{singletL'} 
for  $\HH_{singl}$
can be found in \cite{kausch}, section 3.},
\be \label{singletL'}
\HH_{singl} = {\rm Span} \left\{ \LL_{-k_1}  \dots \LL_{-k_n} 
 c_1 | 0  \ra  
\;, k_i \geq 2 \right\}  \otimes \HH_{matter} \,
 \label{claim}
\ee
where 
\be
\LL_k \equiv  L^{g}_k +k j^{gh}_k + \delta_{k,0} = 
\sum_{n=-\infty}^{+\infty} (k-n) : b_n c_{k-n}: \,.
\label{LL}
\ee
The statement that
\ben
 \label{siegelsingl}
 \ta_{Siegel} &  \in &   \HH^{(1)}_{univ} \cap \HH_{twist +} \cap \HH_{singl} =  \\
&&    {\rm Span} \left\{ \LL_{-k_1}  \dots \LL_{-k_n} 
\, L^m_{-j_1} \dots  L^m_{-j_l}  \,  c_1 | 0  \ra  
\;, k_i \geq 2, j_i \geq 2, \;  { \sum} k_i + \sum j_i \in  2 {\rm \bf N} \right\} \nonumber
\een
summarizes all the known 
{\it linear}  symmetries of the Siegel gauge tachyon condensate.
Other exact constraints (quadratic identities \cite{schnabl}) are
considered in section \ref{quadraticS}.

\subsection{Level-Truncation and Gauge Invariance}
\label{extraS}

We measure the level $L$ of a Fock state
with reference to the zero momentum tachyon $c_1 |0 \ra$,
which we define to be level zero, 
in other terms $L \equiv L_0 +1$. As usual, the level truncation
approximation $(L, N)$ is obtained by truncating the
string field to level $L$, and keeping interactions
terms in the OSFT action up to total level $N$,
with $ 2 L \leq N \leq 3 L$. In our numerical
work we have systematically explored both
the $(L, 2L)$ scheme, which is (naively) the most
efficient, and the $(L,3L)$ scheme, which is the most natural.
In section \ref{L2L} we discuss some empirical
differences between these two schemes.

\smallskip
\begin{table}
\begin{tabular}{|l|l|l|l|l |l|l|l|l |l|l|l|}
\hline
 $  L$  &  0 & 2 & 4 & 6 & 8 & 10 & 12 & 14 & 16 & 18 & 20 \\
\hline
\hline
 $M_{L,1}  $&      1&  2 &  6 &  17 &  43 &  102 & 231 &  496 &  1027 &  2060 & 4010   \\
\hline
$N_L$& 1 &  3 & 9 &  26 & 69 &  171 &  402  & 898 &  1925 &  3985 & 7995 \\
\hline
\hline
$M_{L,2}$ &  0 & 1 &  4 &  12 &  32 &  79 &  182 &  399 &  839 &  1700 & 3342   \\
\hline
$N'_{L}$&   0 & 1 &  5 &  17 &  49 &  128 &  310 &  709 & 1548 &  3248 &  6590 \\
\hline
\end{tabular}
\caption{Dimensions of some relevant subspaces
of $\HH_{univ}$.\label{partitions}
}
\end{table}

The most economic representation of $\ta_{Siegel}$
is using the basis \refb{siegelsingl}, but unfortunately
we have not found a simple algorithm to perform
computations within the SU(1,1) singlet subspace\footnote{The 
twisted ghost Virasoro's 
$L'^g_n$ do not have simple conservation laws on the
cubic vertex.}. We shall work instead
with the universal basis \refb{universal} using fermionic
ghost oscillators.  In this basis,
the number $N_L$ of modes
in $\ta_{Siegel}$ truncated at level $L$ (with $L$ an even integer)
is given by
\be
N_L =  \sum_{j=0}^{L/2}  M_{2j, 1}  \, ,
\ee
where $M_{l, g}$ denotes the number of  
Siegel Fock states in $\HH_{univ}$ with level $l$ and ghost number
$g$.  $M_{l,g}$  which is computed by the generating function
\be
 \sum_{l,g} M_{l,g} x^l y^{g-1} = \prod_{p=2}^\infty \frac{1}{1-x^p} \prod_{q=1}^\infty (1 + x^q y) (1 + \frac{x^q}{y} ) \,.
\ee

The Siegel gauge-fixed eom's \refb{Siegelequ}
truncated at level $L$ are a system of
$N_L$ equations in $N_L$ unknowns. As we discuss
in appendix A, the solution can be found very efficiently
using the Newton method. By construction the resulting string field
$\ta^L_{Siegel}$ solves the truncated
Siegel gauge eom's with extremely good accuracy.
However the full gauge invariant eom's \refb{OSFTeq}
impose an extra set of
constraints on the solution. Recall that
the BRST operator can be written as
\be
Q_B = c_0 L_0 -2 b_0 X + \widetilde Q \label{Q} \, ,
\ee
where
\be
\widetilde Q = \sum_{\stackrel{m,n \neq 0}{m+n \neq 0}} \frac{m-n}{2}\, c_m c_n b_{-m-n}
+ \sum_{n \neq 0} c_{-n} L^{m}_n \,.
\ee
The extra conditions on a Siegel string field are
then
\be \label{extraE}
\widetilde Q \Psi + b_0 c_0 (\Psi * \Psi) = 0\,.
\ee
At level $L$, this equation 
entails $N'_L$ extra constraints
on $\ta_{Siegel}^L$, with
\be
N'_L = \sum_{j = 0}^{L/2}  M_{2 j, 2} \, .
\ee
Table \ref{partitions} shows the numbers 
$M_{L, 1}$, $N_L$, $M_{L,2}$ and $N'_L$ up to $L=20$.

\smallskip

The role of  equation \refb{extraE} is simply to enforce extremality
of the solution along gauge orbits.
However, in principle there could be an issue about  the non-perturbative
validity of the Siegel gauge condition (are gauge orbits 
non-degenerate at the non-perturbative Siegel gauge vacuum? \cite{ET2}).
Moreover,  the level truncation procedure explicitly breaks gauge invariance,
which is formally recovered only as $L \to \infty$.
Thus equation \refb{extraE} gives  an independent set of constraints
which are not a priori satisfied by the level-truncated solution.
If Siegel gauge is a consistent gauge choice and if
gauge invariance is truly recovered in the infinite level limit,
then we expect \refb{extraE} to hold asymptotically as $L \to \infty$.
This is a very non-trivial consistency requirement on $\ta^L_{Siegel}$.
Numerical evidence for this is examined in section \ref{fulleoms}.

\sectiono{The Level-Truncated Tachyon Condensate}

Using the numerical methods outlined in appendix  A,
we have determined $\ta_{Siegel}$
up to $L=18$, both in the $(L,2L)$ and in the $(L,3L)$ schemes.
 $(L,3L)$ results appear to be better behaved (we come back
back to this point in section \ref{L2L}),
and in this section we only consider this scheme. 
It is clearly impossible to reproduce here all the coefficients
of the tachyon condensate up to level $18$.  We give some sample
results in Table \ref{sample}. Our complete numerical data
will be made available at \cite{webpage}.

\smallskip

In this section we perform some consistency checks of the level-truncation results,
verifying some {\it exact} properties that the tachyon condensate
must obey. 

\subsection{SU(1,1) invariance}

We have systematically checked that our solutions
for the tachyon condensate can be written in the basis
\refb{siegelsingl}, and thus obey the full SU(1,1) invariance.
This property holds with perfect accuracy (that is, with the
same precision as the number of significant digits that we keep,
which is 15 for double-precision variables in C++). This is nice, but not surprising,
since the SU(1,1) generators commute with $L_0$, and thus
SU(1,1) is an  exact  symmetry of the level-truncated theory.

 \begin{table}
\begin{tabular}{|l|r|r|r|r|}
\hline
& $L=4$ & $L=6$ & $L=8$ & $L=10$ \\
\hline
$c_{1}|0\ra$ & 0.548399 & 0.547932 & 0.547052 & 0.546260 \\
\hline
$c_{-1}|0\ra$ & 0.205673 & 0.211815 & 0.215025 & 0.216982 \\
\hline
$L^{m}_{-2}c_{1}|0\ra$ & 0.056923 & 0.057143 & 0.057214 & 0.057241 \\
\hline
$c_{-3}|0\ra$ & -0.056210 & -0.057392 & -0.057969 & -0.058290 \\
\hline
$b_{-2}c_{-2}c_{1}|0\ra$ & -0.033107 & 0.034063 & 0.034626 & 0.034982 \\
\hline
$b_{-3}c_{-1}c_{1}|0\ra$ & 0.018737 & 0.019131 & 0.019323 & 0.019430 \\
\hline
$L^{m}_{-2}c_{-1}|0\ra$ & -0.0068607 & -0.0074047 & -0.0076921 & 0.0078698 \\
\hline
$L^{m}_{-4}c_{1}|0\ra$ &  -0.005121 & -0.005109 & -0.005102 & -0.005095 \\
\hline
$L^{m}_{-2}L^{m}_{-2}c_{1}|0\ra$& -0.00058934 & -0.00062206 & -0.00063692 & -0.00064553 \\
\hline
\end{tabular} \vskip .075in
\begin{tabular}{|l|r|r|r|r|}
\hline
 & $L=12$ & $L=14$ & $L=16$ & $L=18$  \\
\hline
$c_{1}|0\ra$ & 0.545608 & 0.545075 & 0.544637 &  0.544272\\
\hline
$c_{-1}|0\ra$ & 0.218296 & 0.219236 & -0.219942 & -0.220491\\
\hline
$L^{m}_{-2}c_{1}|0\ra$ & 0.057252 & 0.057256 & 0.057257 &0.057257 \\
\hline
$c_{-3}|0\ra$ & -0.058489 & -0.058625 & -0.058721 & -0.058794\\
\hline
$b_{-2}c_{-2}c_{1}|0\ra$ & 0.035225 & 0.035402 & 0.035535 & 0.035640 \\
\hline
$b_{-3}c_{-1}c_{1}|0\ra$ & 0.019496 & 0.019542 & 0.019574 & 0.019598 \\
\hline
$L^{m}_{-2}c_{-1}|0\ra$ & 0.0079906 & 0.0080782 & 0.0081445 & 0.0081966\\
\hline
$L^{m}_{-4}c_{1}|0\ra$ & -0.005090 & -0.005086 & -0.005082 & -0.005079\\
\hline
$L^{m}_{-2}L^{m}_{-2}c_{1}|0\ra$ & -0.00065124 & -0.00065532 & -0.00065839 & -0.00066081\\
\hline
\end{tabular} \caption{\label{sample} $(L,3L)$ level-truncation results for the lowest modes
of $\ta_{Siegel}$. }
\end{table}

\subsection{ Out-of-Siegel Equations}
\label{fulleoms}

\begin{table}[h]
\begin{tabular}{|l|l|l|l|}
\hline
&$L=6$&$L=14$&$L=\infty$\\
\hline
\hline
$c_{-2}c_{1}|0\ra$&0.00841347 & 0.00257255  & -0.0000400232\\
\hline
$c_{-4}c_{1}|0\ra$&-0.0103276&-0.00307849  &0.0000536768 \\
\hline
$c_{-1}c_{-2}|0\ra$&0.0107901&0.00483115  &0.000005367 \\
\hline
$L^{m}_{-2}c_{-2}c_{1}|0\ra$&0.000892329 &0.000612637 &0.00000877198 \\
\hline
$L^{m}_{-3}c_{-1}c_{1}|0\ra$& -0.00212947 & -0.000877716 &0.00000163665 \\
\hline
$c_{-6}c_{1}|0\ra$& 0.0130217&0.00341282  &0.0000208782 \\
\hline
$c_{-4}c_{-1}|0\ra$& -0.0110576 &-0.00431119 &-0.000160066 \\
\hline
$c_{-3}c_{-2}|0\ra$& 0.00360400& 0.00160614 &-0.0000134344 \\
\hline
$b_{-2}c_{-3}c_{-1}c_{1}|0\ra$&-0.00306293 &-0.000919219  &-0.0000799493 \\
\hline
$b_{-3}c_{-2}c_{-1}c_{1}|0\ra$&-0.00324329&-0.00114819  &-0.0000488214 \\
\hline
$L^{m}_{-2}c_{-4}c_{1}|0\ra$&0.000132483 &-0.000183042  &-0.0000162206 \\
\hline
$L^{m}_{-2}c_{-2}c_{-1}|0\ra$& -0.00188148&-0.000811710 &-0.0000098375 \\
\hline
$L^{m}_{-3}c_{-3}c_{1}|0\ra$&0.000834397&0.000303847  &-0.0000004570 \\
\hline
$L^{m}_{-4}c_{-2}c_{1}|0\ra$&0.000127107&0.0000135260  &0.0000021124 \\
\hline
$L^{m}_{-2}L^{m}_{-2}c_{-2}c_{1}|0\ra$& -0.000179524&-0.0000980000 &-0.0000014704 \\
\hline
$L^{m}_{-5}c_{-1}c_{1}|0\ra$& 0.000903154&0.000310410 &0.0000131051 \\
\hline
$L^{m}_{-3}L^{m}_{-2}c_{-1}c_{1}|0\ra$&0.000271962&0.000105286  &0.0000014747 \\
\hline
\end{tabular}
\caption{\label{full} Sample $ (L,3L)$ level-truncation results for the 
out-of-Siegel equations of motion.
The table shows  data for $L=6$ and  $L=14$, 
and  $L =\infty$ extrapolations obtained from the data for $2 \leq L \leq 14$
with a polynomial fit in $1/L$.} 
\end{table}

We now turn to  the crucial check of the extra
conditions imposed by the full equations of motion before gauge-fixing.
We were able to carry out this computation up to $L=14$.
Table \ref{full} shows some sample results for the 
string field \refb{extraE} evaluated for $\Psi = \ta_{Siegel}$. 
The extra constraints are satisfied already
very well at $L=6$, and significantly better at $L=14$\footnote{This behavior is common
to the higher level modes not reproduced in Table \ref{full}. 
}.
This is  happening thanks to large
cancellations between the two terms in \refb{extraE}\footnote{
At $L=14$, each term in \refb{extraE} is typically 
one or two orders of magnitude
bigger than their sum.}, as can be easily
checked by applying the operator $\widetilde Q$ to the results in Table \ref{sample}.

Even more remarkable are the extrapolations of the data to $L = \infty$,
which give values two or three orders of magnitude smaller than
the $L=14$ results!  Our extrapolation method 
consists in interpolating the data with a polynomial
in $1/L$ of  {\it maximum} degree
(that is, with as many parameters as the number of
data points).  For example,
for the mode $c_{-4} c_1 | 0 \ra$ we have six data points
($L=4,6,8,10,12,14$) and we use a polynomial in $1/L$ of degree five.
Empirically, this method gives better results
($L=\infty$ extrapolations closer to zero)
than making fits with polynomials in  $1/L$ of lower degree.

This analysis  leaves little doubt that
 the full equations of motion are satisfied as $L \to \infty$.

\subsection{Exact Quadratic Identities}
\label{quadraticS}

\begin{table}
\begin{tabular}{|l|r|r|r|r|r|r|r|r|}
\hline
     & $L^m_{2}$ & $L^m_{4}$ & $L^m_{6}$ & $L^m_{8}$ & $L^m_{10}$ & $L^m_{12}$ & $L^m_{14}$ & $L^m_{16}$\\
\hline
2 & 1.127927 &   &   &   &  &  &  &  \\
\hline
4 & 1.069643 & 1.079864 &   &   &  &  &  &  \\
\hline
6 & 1.046467 & 1.051898 & 1.053517 &   &  &  &  &  \\
\hline
8 & 1.034587 & 1.037554 & 1.040767 & 1.036977 &  &  &  & \\
\hline
10&1.027439 & 1.029304 & 1.031367 & 1.033082 & 1.025628 & &  & \\
\hline
12&1.022688 & 1.023975 & 1.025369 & 1.026797 & 1.027437 & 1.017346 &  & \\
\hline
14&1.019312 & 1.020257 & 1.021261 & 1.022317 & 1.023271 & 1.023102 & 1.011026 & \\
\hline
16& 1.016795 & 1.017520 & 1.018279 & 1.019076 & 1.019875 & 1.020461 & 1.019662 & 1.006039 \
\\
\hline
$\infty$ & 0.999916 & 0.999877 & 1.00429 & 1.00526 &  &  &  & \\
\hline
\end{tabular} \caption{ \label{quadratic}
$(L,3L)$ level-truncation results for Schnabl's
quadratic matter identities. The table shows the 
values for the ratios $R_n$ of  equ. \refb{ratios}.} 
\end{table}

As pointed out by Schnabl \cite{schnabl},  
any solution of the OSFT  eom's must obey certain exact quadratic identities
that follow from the existence of anomalous
derivations of the star product.  An infinite set of  identities
is obtained from the anomalous derivations $K^m_{2n} = L^{m}_{2n}-L^{m}_{-2n}$. They are
\cite{schnabl}:
\be\label{quadraticE}
\la \Psi | c_{0}L^m_{2n}
|\Psi\ra  =    (-1)^n \, \frac{65}{54} \,   \, \la \Psi | c_{0}L_0 |\Psi\ra \, ,
\ee
where $\Psi$ is a solution in Siegel gauge.

\smallskip

In Table \ref{quadratic} we show the level-truncation results for the ratios
\be \label{ratios}
R_n \equiv(-1)^n \, \frac{54}{65}
\frac{ \la \ta | c_{0}L^m_{2n} |\ta \ra}{  \la \ta | c_{0}L_0 |\ta \ra } \, ,
\ee
which are of course predicted to be exactly one.
The quadratic identities are satisfied quite well
already at low levels, and the extrapolations to $L = \infty$ 
(performed again with  polynomials
in $1/L$ of maximum degree) give really good results. 

\smallskip

Both the quadratic identities just analyzed and the 
out-of-gauge eom's \refb{extraE} are exact constraints on the 
solution that are broken by level-truncation. We have found 
 that the level-truncated answers for this class of observables are 
very accurately converging to their known exact values as $L \to \infty$. 
This is strong evidence
for  the idea that level-truncation is a convergent approximation scheme.
We have also learnt that maximal polynomials  
in $1/L$ give very precise extrapolations.  
It seems safe to assume that this should be a universal feature,
and in the following we shall adopt the same
extrapolation technique to quantities  whose exact 
asymptotic value is  a priori unknown\footnote{Polynomials
in $1/L$ have also been used in the extrapolation procedure of
of \cite{wati}.  It was also noted  in \cite{perturbative}
that large level results appear to have corrections of order $1/L$, 
although there the definition of level is somewhat different.}.

\section{Extrapolations to Higher Levels}

Encouraged by the successful extrapolations to $L = \infty$ described
in the previous section, and inspired by Taylor's analysis \cite{wati},
we have set up a systematic scheme to extrapolate
to higher levels the results for the vacuum energy and 
for the  tachyon condensate string field. In this section
we focus on the results for the vacuum energy, while
in the next we shall examine the results for the 
individual coefficients of the tachyon condensate.

\smallskip

Unless explicitly stated otherwise,  
the use of the $(L,3L)$ scheme is implied in the rest of the paper.
We  justify this choice in section \ref{L2L}, where we briefly
contrast $(L,2L)$  versus $(L,3L)$ results. 

\subsection{Extrapolations of the Tachyon Effective Action  }

The basic idea of our method has already been explained
in the introduction.  The first step is the computation of
the tachyon effective action  $V_L(T)$, obtained
 by integrating out the higher modes up to level $L$. 
Some details of how this is done numerically are
explained in  appendix A.
Figure \ref{effectiveA} shows the plots of $V_L(T)$ for $0 \leq L \leq 16$.
There is good convergence as $L$ increases, indeed the 
curves for $L \geq 6$ are  indistinguishable 
on the scale of Figure \ref{effactions}.
\begin{figure}[t]
\label{effectiveA}
\begin{center}
\epsfysize=8cm
\epsfbox{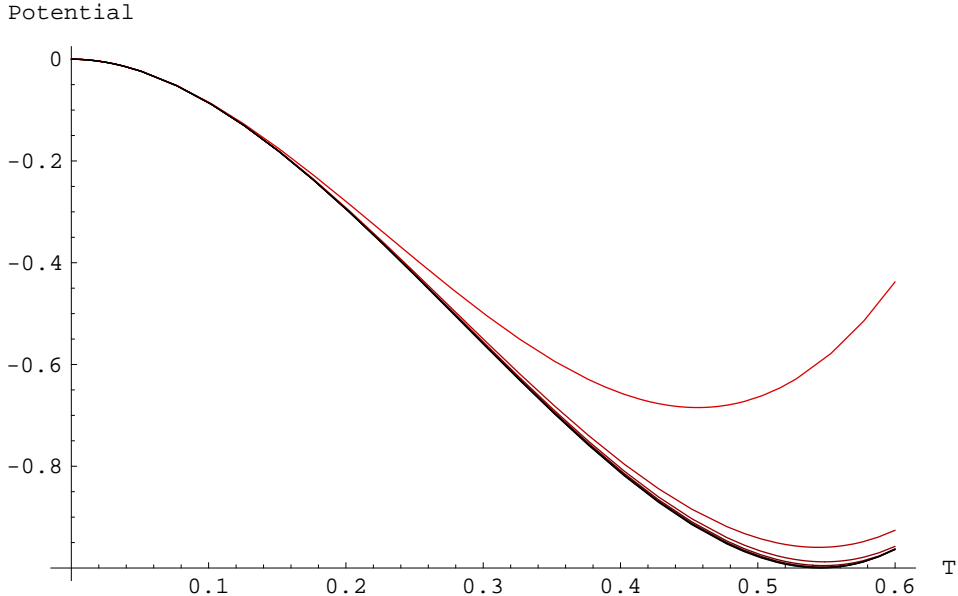}
\end{center}
\caption{Plots of the tachyon effective potential $V_L(T)$
at level $L$,  for $L$ between zero (uppermost curve) 
and 16 (lowermost curve). The curves for $L=6,10,12,14,16$ appear
superimposed in the figure.}
\label{effactions} \end{figure}

\smallskip

For our extrapolations, we focus on a interval for  the tachyon vev
around the non-perturbative vacuum. We take $0.54 \leq T \leq 0.55$.
The function $V^{(M)}_L(T)$, where $M$ is an even integer $\leq 16$,
is  the extrapolation  `of order $M$'  of the tachyon effective
action at level $L$, and  is constructed as follows.
We fix the dependence on $L$   by writing
\be   \label{VMdef}
V^{(M)}_L (T) = \sum_{n=0}^{M/2 +1}  \frac{a_n (T)}{(L+1)^n} \,  ,
\ee
for some coefficients functions $a_n(T)$. The
functions $a_n (T)$ 
 are determined by imposing the  conditions
\be
V^{(M)}_L (T) = V_L (T)\, ,   \qquad \; {\rm for \;}  L =0,2, \cdots, M  \,,   \quad \quad \forall \; T\in [ 0.54, 0.55] \,.
\ee
 In other terms, we interpolate the $M/2 +1$ values
$\{V_L(T) | L=0,2,\dots M \}$ with a polynomial in $1/(L+1)$ 
that passes  through all the data points\footnote{The
 rationale
for using  polynomials in $1/(L+1)$
 rather than $1/L$ is that
we wish to include also the data for $L =0$. This works
 somewhat better than excluding
the $L = 0$ point and making extrapolations in $1/L$. 
Committed readers can find more about this technicality in footnote
\ref{technicality}.}.

\begin{figure}[ht]
\begin{center}
\epsfysize=9cm
\epsfbox{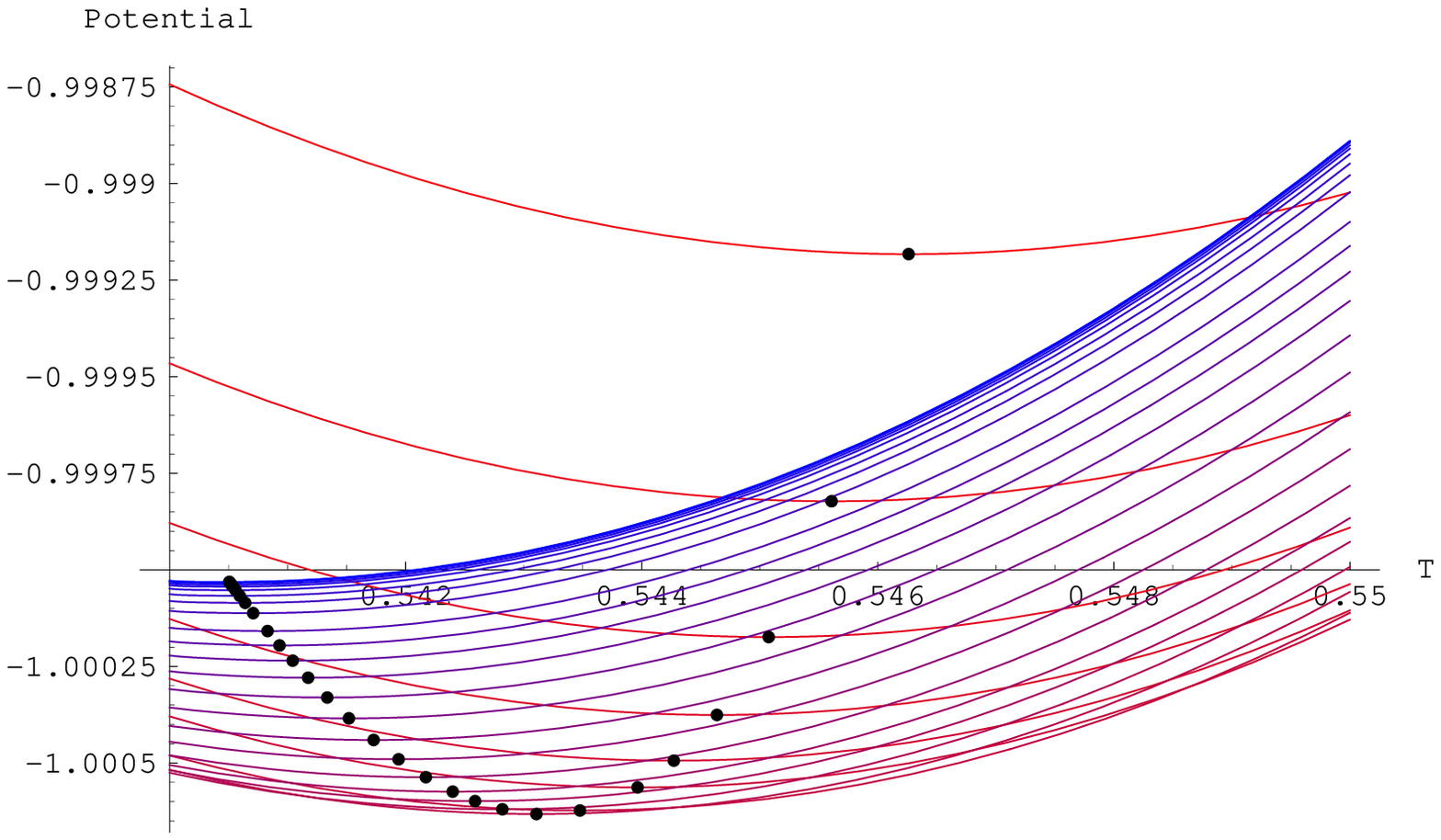}
\end{center}
\caption{Plots of the order 16 estimates
 $V_L^{(16)}(T)$ for the  effective tachyon potential,  for 
some sample values of $L \geq 10$. The minimum of each curve is indicated by
a black dot, which by definition has coordinates $(T^{(16)}_L, E_L^{(16)})$.
The isolated uppermost plot corresponds to $L=10$.  To follow
the curves from $L=10$ to $L=\infty$, focus on the position of the
minima: as $L$ increases, the dot moves from right to left
({\it i.e.}, the tachyon vev decreases). As $L \to \infty$, the curves
crowd towards an asymptotic function with minimum at $
(T_\infty^{(16)}, E_{\infty}^{(16)}) =(0.5405, -1.00003)$.}
\label{effminima} 
\end{figure}

\smallskip
\begin{figure}[ht]
\leavevmode
\begin{center}
\epsfysize=8.5cm
\epsfbox{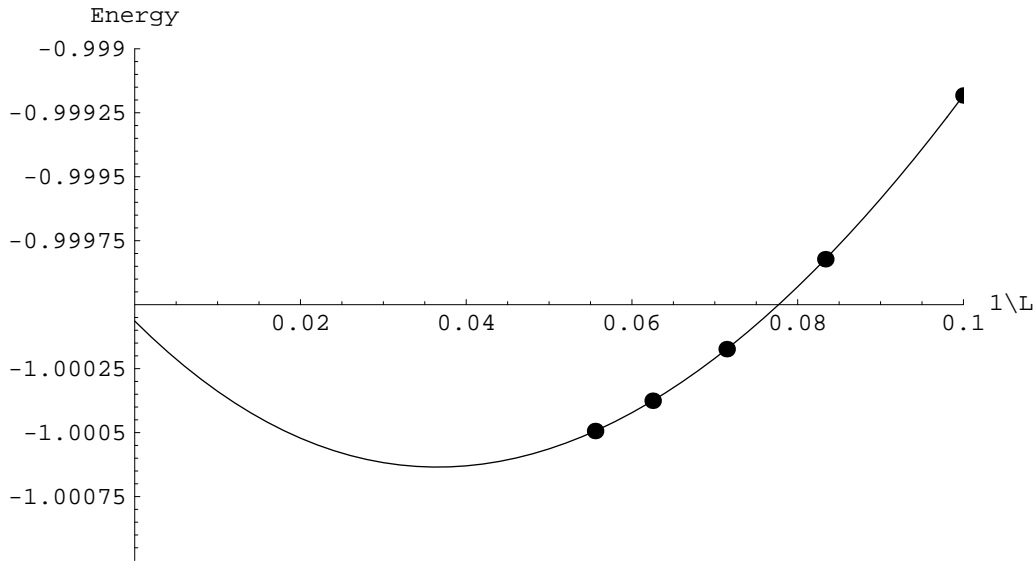}
\end{center}
\caption{Plot of $E_L^{(16)}$
as a function of $1/L$. The black dots
represent the exact values up to $L=18$
computed by direct level-truncation (Table 1, $(L,3L)$ scheme).
To first approximation, the curve in the figure is roughly a parabola: since
the energy overshoots -1 at $1/L =1/14 \simeq 0.07 $, 
we have a visual understanding of the position of the minimum
of the energy around $1/L =(1/14)/2 = 1/28  \simeq 0.036$.}
\label{refinedaction} 
\end{figure}

\begin{table}[ht]
\begin{tabular}{|l|l|l|l|l|}
\hline
&$L=4$&$L=6$&$L=8$&$L=10$ \\
\hline
$M=4$&-0.98782176&-0.99546179&-0.99850722&-1.0000023 \\
\hline
$M=6$&           &-0.99517712&-0.99798495&-0.99930406 \\
\hline
$M=8$&           &           &-0.99793018&-0.99918359 \\
\hline
$M=10$&          &           &           &-0.99918246 \\
\hline
\hline
&$L=12$&$L=14$&$L=16$&$L=18$ \\
\hline
$M=4$&-1.0008372&-1.0013461&-1.0016765&-1.0019017 \\
\hline
$M=6$&-1.0000079&-1.0004169&-1.0006692&-1.0008317 \\
\hline
$M=8$&-0.99982545&-1.0001796&-1.0003843&-1.0005057 \\
\hline
$M=10$&-0.99982266&-1.0001750&-1.0003780&-1.0004979 \\
\hline
$M=12$&-0.99982226&-1.0001739&-1.0003759&-1.0004947 \\
\hline
$M=14$&           &-1.0001737&-1.0003755&-1.0004938 \\
\hline
$M=16$&           &          &-1.0003755&-1.0004937 \\
\hline
$M=18$&           &          &          &-1.0004937 \\
\hline
\end{tabular}
\caption{
Estimates $E^{(M)}_L$ for the vacuum energy
obtained from extrapolations of the effective tachyon  potential,
at various orders  $M$ and for $L \leq 18$.
Data in the $(L,3L)$ scheme. By
definition, the diagonal entries $E^{(M=L)}_L$ coincide with the exact
computation from direct level-truncation at level $(L, 3L)$ (Table 1).
\label{Maction3L}  }
\end{table}

Our best estimate for the tachyon effective action at level $L$
is  the function $V_L^{(16)}(T)$. Figure \ref{effminima} shows the 
plots of $V_L^{(16)}(T)$  for $L$ between ten and infinity.
The position of the minimum in each curve defines
our order $M=16$ estimates $T^{(16)}_L$  and $E^{(16)}_L$ for
the tachyon vev and vacuum energy at level $L$.
We can follow very clearly the behavior of the minima
as $L$ increases.  The energy falls below -1, reaches
its lowest point in   $L =28$ curve, and then turns back to 
approach asymptotically  the value $E^{(16)}_\infty = -1.00003$ !
In Figure \ref{refinedaction} we see the same phenomenon in  a plot
of $E^{(16)}_L$ as a function of $1/L$. 

\smallskip

It is interesting to consider how the extrapolations 
change as we vary $M$. 
Table \ref{Maction3L} shows the estimates $E^{(M)}_L$ 
up to $L=18$, while Table \ref{Mtachyon3L} (appendix B)
shows the analogous estimates for the tachyon vev.
By construction, the diagonal entries  $E^{(L)}_L$ and $T^{(L)}_L$
 are simply the exact values  obtained by direct level-truncation at level $L$. 
One can observe from the tables
that the method has remarkable predictive power. 
For example, by only knowing level-truncation results up
to level 10, one can obtain the prediction $E^{(10)}_{16}=-1.0003780$
for the energy at level 16, to be compared to the exact
value $E_{16} = -1.0003755$.
 We thus feel  quite comfortable in trusting the extrapolations even for $L$ large.  
Figure \ref{logactions} and Table \ref{Einf} (already discussed in the introduction)
illustrate the main features of the larger $L$ results for the vacuum energy,
for various $M$'s.
It is pleasant to observe  that, as $M$ increases,
the estimates $L^{(M)}_{min}$ and $E^{(M)}_{min}$
(Table \ref{Einf}) quickly  reach stable values, while $E^{(M)}_{\infty}$ 
steadily approaches minus one\footnote{
\label{technicality}
Finally we would like to comment on how results change if instead of using a polynomial
extrapolation in $1/(L+1)$ we use an extrapolation
in $1/L$ (excluding the $L=0$ point), or  alternatively
we keep the $L=0$ point and use a polynomial in $1/(L+a)$
for some other $a > 0$.
One finds that for $M=16$ the differences among all these schemes
are very minor, even for a wide range of reasonable values of $a$ (say $0.1 < a < 3$).
For $M < 10$, including the $L =0$ data (and using $1/(L+a)$)
 works somewhat better than excluding it (and using $1/L$).
For example the prediction $E_{16}^{(8)}$ 
 obtained excluding $L=0$ differs by the exact value 
by an error of $0.0003$, which is 30 times bigger 
than for the prediction obtained including $L=0$.
 All of this scheme-dependence is expected to disappear for $M$ large,
and  indeed is already irrelevant at $M=16$.}.

\smallskip

It is remarkable that extrapolations to higher levels
work so well. The data have a smooth and predictable dependence
on $L$, which is very well captured by polynomials in $1/L$. 
This property was  not  {\it a priori} obvious,
and indeed it appears to be true only for $(L,3L)$ data,
as we shall see in section \ref{L2L}.

\subsection{Comparison with Straightforward Extrapolations}

The method just described appears to work remarkably well.
To which extent does the success of the method depend
on the sophisticated 
idea of   extrapolating  the {\it functional form}
of the tachyon effective action?  We can answer
this question by considering the more 
straightforward
procedure of simply extrapolating the values
of the vacuum energies $E_L$ (as opposed to the full
functions $V_L(T)$).    We define the `straightforward'
order $M$ estimate ${\widetilde E}^{(M)}_L$ at level $L$
by considering the data $\{ E_L  |  0 \leq L \leq M \}$,
and interpolating them with a polynomial in $1/(L+1)$
 of maximum degree. (This is in complete analogy with \refb{VMdef}).
 The results for $L \leq 18$ in the $(L,3L)$ scheme
are presented in Table \ref{Mnaive}, while in Table \ref{L2Lnaive} we
give the extrapolations to $L = \infty$. 
 We see that for $M <10$, the more sophisticated method 
gives much more accurate
predictions 
(compare Table~\ref{Maction3L} and Table~\ref{Mnaive}:
 for example $E^{(8)}_{18}$, ${\widetilde E}^{(8)}_{18}$ and
the exact value $E_{18}^{(18)}$). 
However for $M >10$ there is no significant difference between the two procedures\footnote{
A comparison with the results
of \cite{wati}, which in our language correspond to $M =10$,
shows that the accuracy of the perturbative method of \cite{wati}
seems comparable with the accuracy of the straightforward extrapolation. 
For $M=10$ our non-perturbative method based on the tachyon effective action
appears to be more accurate.}.  We also 
compared the results
for the individual coefficients of the tachyon string field
obtained with the two procedures, and found a very similar pattern.
\begin{table}[h]
\begin{tabular}{|l|l|l|l|l|}
\hline
&$L=4$&$L=6$&$L=8$&$L=10$ \\
\hline
$M=4$&-0.98782176&-0.99730348&-1.0020443&-1.0048888 \\
\hline
$M=6$&           &-0.99517712&-0.99845611&-1.0002959 \\
\hline
$M=8$&           &           &-0.99793018&-0.99921882 \\
\hline
$M=10$&          &           &           &-0.99918246 \\
\hline
\hline
&$L=12$&$L=14$&$L=16$&$L=18$ \\
\hline
$M=4$&-1.0067851&-1.0081397&-1.0091556      &-1.0099457 \\
\hline
$M=6$&-1.0014693&-1.0022814&-1.0028762      &-1.0033304 \\
\hline
$M=8$&-0.99991100&-1.0003187&-1.0005753    &-1.0007448 \\
\hline
$M=10$&-0.99982332&-1.0001767&-1.0003811  &-1.0005023 \\
\hline
$M=12$&-0.99982226&-1.0001738&-1.0003758  &-1.0004946 \\
\hline
$M=14$&           &-1.0001737&-1.0003755              &-1.0004939 \\
\hline
$M=16$&           &          &-1.0003755                         &-1.0004937 \\
\hline
$M=18$&           &          &                                              &-1.0004937 \\
\hline
\end{tabular}
\caption{\label{Mnaive}
The estimates
${\widetilde E}^{(M)}_L$ for the vacuum energy in
the $(L,3L)$  scheme,  obtained with the 
`straightforward'  polynomial extrapolation in $1/(L+1)$. }
\end{table}

\smallskip

We conclude that with  the sophisticated procedure
one can achieve  remarkable accuracy even for small
$M$, where a naive extrapolation would work quite poorly. 
However if one is willing to perform level-truncation
up to level 12 or above, the simpler extrapolation procedure
is equally effective.

\subsection{ $(L, 3L)$ versus $(L, 2L)$}
\label{L2L}

\begin{table}[ht]
\begin{tabular}{|l|l|l|l|}
\hline
     &$(L,3L)$&$(L,2L)$ \\
\hline
$M=6$&-0.998698&-0.988625 \\
\hline
$M=8$&-0.999784&-1.00261 \\
\hline
$M=10$&-1.00010&-0.999316 \\
\hline
$M=12$&-1.00008&-1.00048 \\
\hline
$M=14$&-1.00004&-0.999655 \\
\hline
$M=16$&-1.00003&-1.00057 \\ 
\hline
\end{tabular}
\caption{\label{L2Lnaive}
The estimates ${\widetilde E}^{(M)}_\infty$ for the asymptotic vacuum energy in
the $(L,3L)$ and $(L,2L)$ schemes,  obtained with the 
`straightforward' polynomial extrapolation in $1/(L+1)$. }
\end{table}
\begin{figure}[ht]
\begin{center}
\epsfysize=8cm
\epsfbox{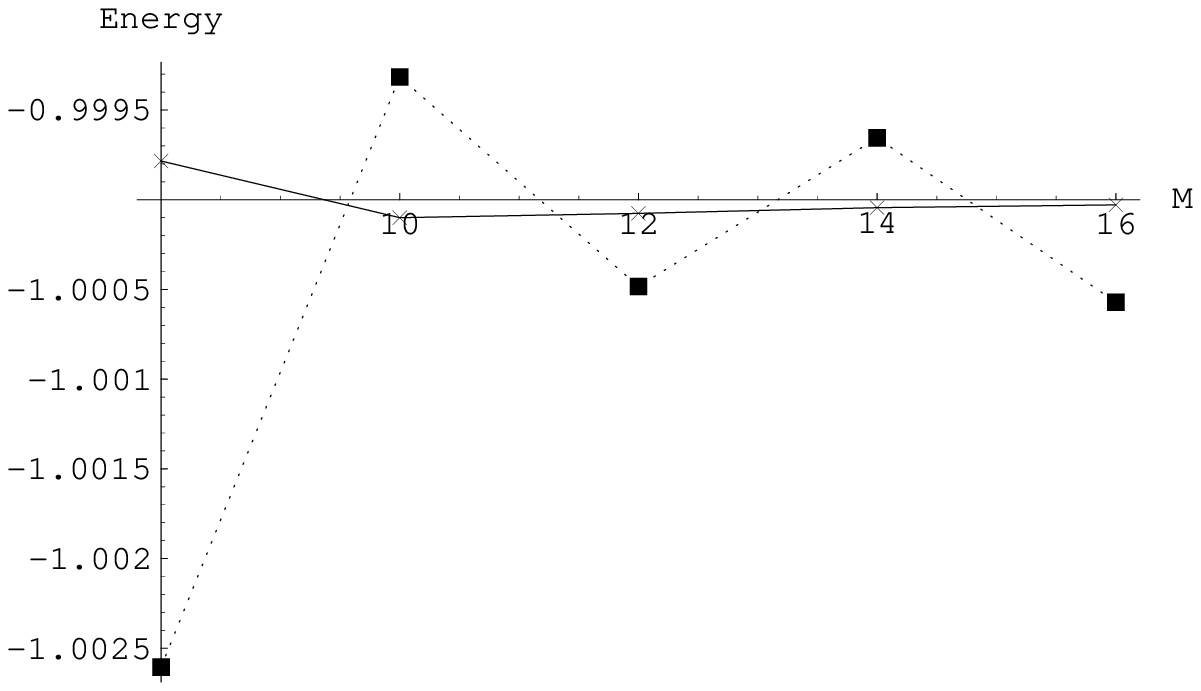}
\end{center}
\caption{
\label{L2Lfig}
Plot of the results in Table \ref{L2Lnaive}. The  continuous line represents
the $(L,3L)$ results, while the dashed line represents the $(L,2L)$ results.
}
 \end{figure}

All the extrapolations described so far are for
results in the $(L,3L)$ scheme. We have investigated
the data in the $(L,2L)$ scheme and concluded
that their behavior as a function of $L$ is not nearly
as smooth: as a consequence, extrapolations
to higher levels are less reliable. 
 A glance at Table \ref{L2Lnaive} and Figure \ref{L2Lfig}
is sufficient to illustrate our point.  We are comparing
the `straightforward' extrapolations 
 of the vacuum energy to $L = \infty$, for various $M$'s,
 obtained with data in the $(L,3L)$ scheme, with the analogous quantities
in the $(L,2L)$ scheme. While the $(L,3L)$ data
have  a really smooth dependence on $M$ 
and converge nicely to  -1, the $(L,2L)$ data have a much more
irregular behavior. A similar pattern is observed
for extrapolations at finite $L$: 
estimates ${\widetilde E}^{(M)}_L$ of exact
results at level $L \leq 18$ are not nearly as
accurate in the $(L,2L)$ scheme as they are in the $(L,3L)$ scheme.
An analogous behavior is found in comparing $(L,3L)$ and $(L,2L)$
 data for the tachyon vev.  
 We have also repeated for $(L,2L)$ data
the full analysis based on extrapolations of
the tachyon effective action, and found no improvement with respect to the
straightforward extrapolations shown in Table~\ref{L2Lnaive}.

\smallskip

It would be interesting to explain these findings from 
an analytic point of view. The $(L,3L)$ scheme can be understood as a cut-off
procedure in which only the kinetic term of the OSFT action is changed,
such to give an infinite mass to  modes with level higher than $L$.
 On the other hand,  in the $(L,2L)$ scheme   both the kinetic and the cubic term of the action
 are changed. This
may explain why $(L,3L)$ data have a simpler dependence
on $L$.

\section{Patterns of the Siegel Gauge Tachyon Condensate} \label{s3}

The individual coefficients of the tachyon condensate can also
be extrapolated to higher levels. The more sophisticated method that we use
is the following: We solve the classical
equations of motion at level $L$ and express all higher modes
$\widetilde \Psi_L$ as functions of the tachyon vev $T$ (see \refb{psitilde}).
We then perform an extrapolation of these
functions of $T$ using  a polynomial in $1/L$ of maximum degree:
this defines in the usual way 
the order $M$ estimates $\widetilde \Psi^{(M)}_L[T]$. Finally by
setting $T = T_L$ we obtain extrapolations for the full tachyon
string field.  Tables \ref{Mu3L} and \ref{Mv3L} (appendix B) 
contain the extrapolations to $L \leq 18$ of the first two higher  modes.
In Table \ref{asymptotic} we give the $M=16$ results
for the extrapolations to $L = \infty$ 
obtained using this method\footnote{The straightforward
method of simply extrapolating the coefficients
with a maximal polynomial in $1/L$ is also possible:
for $M=16$ the difference with the data in Table \ref{asymptotic}
is very minor,  at most or  two units in the last significant digit. }.

\smallskip

We are finally in the position to  look for analytic
patterns in the tachyon condensate, in particular checking
in a more reliable way
the patterns observed in \cite{patterns}. This was  one of the motivations of
our work.

\bigskip

{\it \noindent The Tachyon Vev}

\smallskip
 \begin{figure}[t]
\begin{center}
\epsfysize=9cm
\epsfbox{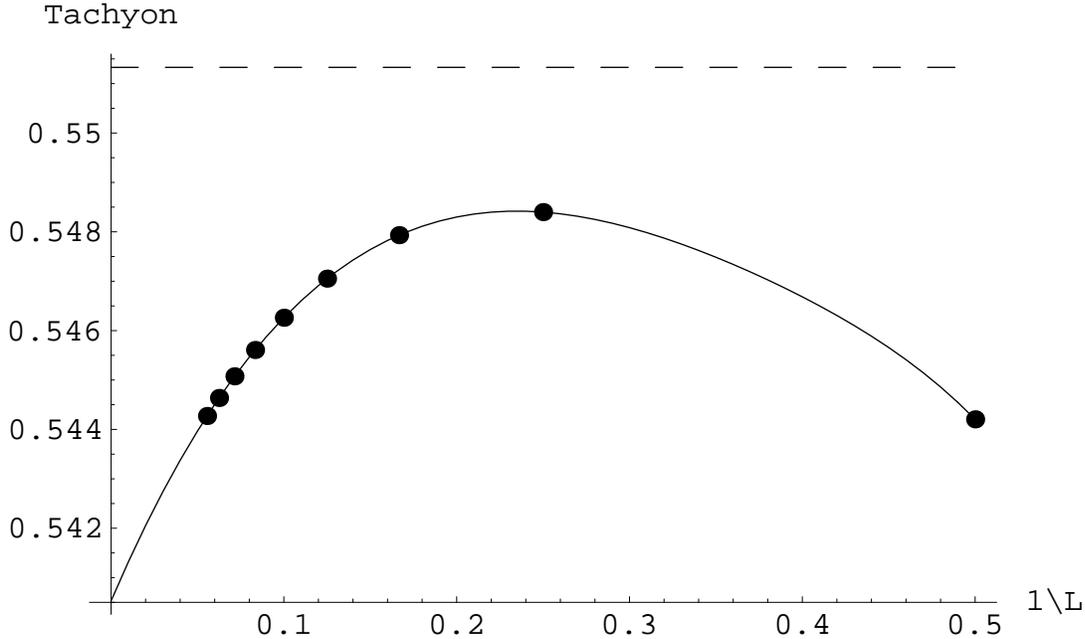}
\end{center}
\caption{Plot of the estimate $T^{(16)}_L$ for tachyon vev
as a function of $1/L$. The dots represent the exact
values $T_L$ for $L =2 \dots 18$.
The dashed line represents the
value $T_{HS} = \sqrt{3}/\pi$. }
\label{tachyon} \end{figure}

Figure \ref{tachyon} is a plot of the estimate for the tachyon vev as a function of $1/L$.
The asymptotic value is $T^{(16)}_\infty \simeq0.5405$.
Clearly the conjecture \cite{HS} for an exact value $T_{HS} = \sqrt{3}/\pi$
is falsified\footnote{This conclusion was  already believed based
on level-truncation results to level 10 \cite{MT}, but at  the time
of the Strings 2002 conference \cite{strings2002}
 we had proposed that somehow
this conjecture could be rescued.
In an attempt to explain the puzzling overshooting of the vacuum energy
in Table 1, we had suggested an {\it ad hoc} renormalization
of the tachyon condensate by an overall multiplicative factor,
such that the tachyon vev is exactly $T_{HS}$. With this
renormalization,
the vacuum energy at $L=18$ becomes very accurately -1. In view
of the new results in \cite{wati} and in this paper, clearly
there is no need for any such mechanism.}. 

\bigskip

\break

{\it  \noindent Universal Ghost Coefficients}

\begin{table}[t]
\begin{center}\def\st{\vrule height 3ex width 0ex}
\begin{tabular}{|l|l|l|l|l|l|l|} \hline
$L$ & $r_{1,1}$ & $r_{3,1}$ & $ r_{5,1}$  &
$r_{3,3}$ & $r_{7,1}$ & $r_{5,3}$
\st\\[1ex]
\hline
\hline
2 & 0.349483 &            &  &  &  &
\st\\[1ex]
\hline
4 & 0.375042 &  -0.102499 &  &  &  &
\st\\[1ex]
\hline
6 & 0.386571  & -0.104743 & 0.0547758  & 0.0208371  &  &
\st\\[1ex]
\hline
8 & 0.393062 & -0.105966 & 0.05544 & 0.0209129  & 0.0358037 & 0.014195
\st\\[1ex]
\hline
10 & 0.397214 & -0.106707 & 0.0558499 & 0.0209927 & 0.0361037 & 0.0142012
\st\\[1ex]
\hline
12 & 0.400096 &  -0.107201 & 0.0561103 & 0.0210499 & 0.0363031 & 0.0142291
\st\\[1ex]
\hline
14 & 0.402212  & -0.107553 & 0.0562894  & 0.0210917 & 0.0364338 & 0.0142521
\st\\[1ex]
\hline
16 & 0.403832 & -0.107818 & 0.0564199 & 0.0211232 & 0.0365254 & 0.0142701
\st\\[1ex]
\hline
18 & 0.405111 & -0.108023 & 0.0565194 & 0.0211478 & 0.0365932 & 0.0142843
\st\\[1ex]
\hline \hline
$\infty$ &0.4160 &-0.1097  &0.05728  &0.02135 &0.03739 & .01456
\st\\[1ex]
\hline \hline
$conj$ &0.407407 &-0.109739  &0.0577148  &0.021328 & 0.037483 & .01439
\st\\[1ex]
\hline
\end{tabular}
\end{center}
\caption{$(L,3L)$ numerical results for the pattern coefficients $r_{n,m}$ 
for the tachyon condensate string field, and their conjectured values.
The $L = \infty$ results are obtained
from the $M=16$ extrapolation procedure based on the effective tachyon potential.} 
\label{ghostpattern} \end{table}

\smallskip

The most accurate pattern discovered in \cite{patterns}
is the remarkable universality of certain ghost coefficients.
If one normalizes the tachyon mode to one, then the coefficients
$r_{n,m}$ of the
modes $c_{-n} b_{-m} c_1 | 0 \ra$ (with $n$ and $m$ {\it odd})
appear to be the same for all known OSFT solutions. Assuming that this is the case,
an analytic prediction for these coefficients
is possible by looking at infinitesimal OSFT solutions
corresponding to exactly marginal deformations of the BCFT.

\smallskip

We reproduce our results for these coefficients for the tachyon condensate
in Table \ref{ghostpattern}  and in Table \ref{higherR} (this
latter table is in appendix B). 
There is certainly a  striking pattern. In particular,
the results of Table \ref{higherR} show that the pattern
persists in the higher-level modes that were not explored in \cite{patterns}.
 However the value
for $r_{1,1}$  is puzzling. This result has not improved with respect to the $L=12$ data
already available in \cite{patterns}\footnote{
In \cite{patterns} we performed an extrapolation
of the data up to $L=12$ using a fit $a + b/L$, and found
$r_{1,1} =  0.411545 $. If one instead applies to the data up to $L=12$
an extrapolation with a maximal polynomial in $1/L$
(as advocated in this paper), one finds $r_{1,1} = 0.415947$.
 Changing extrapolation
scheme or adding more levels does not seem to help in getting
closer to the conjectured answer.}, and the $ 2\%$  difference
from the conjectured value would seem like a real one\footnote{Some of
the higher-level coefficients (Table \ref{higherR})
 also have errors of a few percents, but this need not be a problem, since at higher levels
we expect larger errors.}.

\smallskip

Although
we hesitate to assign a precise error to our extrapolations to $L = \infty$,
it seems 
that this error should be smaller than than $2\%$.  Indeed
all the properties that we know for sure to be exact
(the quadratic identities, the out-of-Siegel equations
and of course the value of the vacuum
energy), are obeyed with an accuracy of  order $10^{-5}$
in the $M=16$ extrapolations\footnote{While the energy is stationary
and so it is affected quadratically by a small change in the coefficients,
Schnabl's identities and the out-of-Siegel equations vary linearly.
It would take some conspiracy
for  the vev of of $c_{-1} | 0 \ra$ to have a $2\%$ error, and at the same
time the corresponding out-of-Siegel equation be so well obeyed
(see the first line in Table \ref{full}, which is linearly influenced by
a small variation of this coefficient).}.

\begin{table}
\begin{tabular}{|r|r|r|r|}
\hline
& ${\cal T}_{Siegel} $&$\Phi \; \;  \; \; $ \\
\hline
$L_{-2}$&.1058&0.1069 \\
\hline
$L_{-4}$&-.009343&-0.009476 \\
\hline
$L_{-2}L_{-2}$&-.001260&-0.001221 \\
\hline
$L_{-6}$&.002648&0.002691 \\
\hline
$L_{-3}L_{-3}$&.0000135&0.0000143 \\
\hline
$L_{-4}L_{-2}$&.000575&0.000594 \\
\hline
$L_{-2}L_{-2}L_{-2}$&-.000009&-0.000016 \\ 
\hline
\end{tabular}
\caption{\label{gnzero}    The left column shows
the normalized matter coefficients of the tachyon condensate
string field for $L = \infty$ (from Table \ref{asymptotic}). The right columns
shows the corresponding coefficients for the
ghost number zero string field,
obtained from an extrapolation to $L = \infty$ of data up to $L =14$,
with a maximal polynomial in $1/L$.  }
\end{table}

\bigskip

{\it \noindent  $L_0\, {\cal T}$ factorization}

\smallskip

In \cite{patterns} it was observed that
the string field $L_0\, {\ta_{Siegel}}$ is
approximately factored  into  a matter times a ghost component.
 In view of our more precise
data, we definitely conclude that this is just a rough pattern,
as already believed in \cite{patterns}. Indeed
the pattern is seen to get
 worse in the $L = \infty$ extrapolations.
Let us give a couple of examples.
 Assuming factorization, 
the normalized coefficient of $L^{m}_{-2} c_{-1} | 0 \ra$
can be obtained by multiplying the two normalized coefficients
of $L^m_{-2}| 0 \ra$ and $ c_{-1} | 0 \ra$. In \cite{patterns}, using the
numerical values at level $(10,30)$, this procedure
gave a prediction with a $4 \%$ error;
  with the $L = \infty$ data in Table \ref{asymptotic},
the error is increased to $8 \%$. Similarly, applying the same
procedure  to $L^m_{-4} c_{-1} | 0 \ra$, one finds
only a  $0.1 \%$ error at level $(10,30)$, but a $1 \%$ error
using the values in Table \ref{asymptotic}.

\bigskip

{\it \noindent  Correspondence with  Ghost Number Zero}

\smallskip

The last pattern noticed in \cite{patterns} is a surprising
coincidence between the OSFT equation and the 
equation
\be \label{zero}
(L_0 -1) \Phi + \Phi * \Phi = 0 \, ,
\ee
where $\Phi$ is a ghost number zero field. 
 This equation can be solved in the universal
space spanned by {\it total} Virasoro's  $L_{-n}$
acting on the vacuum. It was found that the coefficients of the terms
$L_{-k_1} \cdots L_{-k_n}  | 0 \ra$ in the solution of
\refb{zero} are strikingly close to the normalized
coefficients of the matter modes  $L^m_{-k_1} \cdots L^m_{-k_n} c_1 | 0\ra$
of $\ta_{Siegel}$. However the pattern
is somewhat irregular and does not appear
to systematically improve with level. This is confirmed
by our more precise data, a sample of which is presented
in Table \ref{gnzero}.  There is no clear improvement with respect
to the data in \cite{patterns}, and we confirm the conclusion
that this is likely only a `quasi-pattern'.

\bigskip

All in all,  the patterns of \cite{patterns} are
still present in our more precise data, 
and one cannot help the feeling that they  
hint at some analytic clue. However we  seem to conclude
that all these patterns are not exact properties,
with the possible exception of the universality conjecture
for certain ghost coefficients. This last conjecture is generally
 well obeyed, but we found a disturbing discrepancy for the $c_{-1} | 0 \ra$
mode.

\section{Concluding Remarks}

Our results support the idea
that level-truncation  is a completely reliable
approximation scheme for OSFT, with
a convergent limit as the level is sent to infinity.
 All available exact predictions (notably
the value of the vacuum energy)
are accurately confirmed by the data. No inconsistencies
seem to arise from the fact that gauge-invariance is broken at finite level,
 indeed we found strong evidence that it is smoothly restored as 
$ L \to \infty$.  
Quantities computed in level-truncation exhibit a predictable dependence
on level which is very well approximated by polynomials in $1/L$
(at least for the $(L,3L)$ scheme). This 
 allows reliable extrapolations to higher levels.
Combining this observation with 
efficient computer algorithms based on conservation
laws \cite{0006240}, we 
have developed very powerful numerical tools to study OSFT.

\smallskip

In this paper we have focused on the universal subspace
and obtained accurate data for the tachyon condensate. 
An obvious direction for further research is
to use our results to learn about the kinetic term around the tachyon vacuum.
The nature of this kinetic term is still rather mysterious. 
No perturbative open string states are expected to be present,
and numerical evidence for this has already been obtained
\cite{ET1}. Our data will allow a more precise analysis,
and hopefully give new analytic clues.

\smallskip

The most intriguing aspects of the non-perturbative
vacuum are related to the elusive closed string states.
In OSFT, amplitudes for external closed
strings (on a surface with a least one boundary)
are given by correlation functions of certain gauge-invariant open string functionals \cite{ST,
Z,HI, vsft2}. It would be very interesting to compute such amplitudes
in the non-perturbative vacuum. This
should shed some light on the mechanism by which
open string moduli are frozen in the tachyon vacuum,  but closed string moduli are still present.
There are promising ideas for how this may come
about \cite{vsft2, shata, drukker}, but the actual mechanism realized in OSFT is still 
unknown.

\smallskip

Another avenue for future work is 
the application of our methods to more general classical solutions.
It will be straightforward to extend our algorithms
to include the matter states necessary to construct non-universal  solutions,
 {\it e.g.}  tachyon lump solutions \cite{0005036}
and Wilson line solutions \cite{0007153}.  It would also be
very nice to  investigate numerically  time-dependent solutions, 
and demonstrate the existence of 
tachyon matter  \cite{tachyonmatter} in OSFT.  The study of several classical solutions
will help to build the intuition that is needed for analytic progress.

\smallskip

We hope that the  precise data presented in this paper
will encourage other physicists to 
think about analytic approaches to OSFT. 
Our analysis of   the `patterns' of the tachyon condensate  
 throws some doubt even on the  most robust conjecture proposed in \cite{patterns}, although it
does not rule it out completely. 
The search for analytic clues in OSFT solutions is still very open.  
Yet we feel that our  numbers for the tachyon string field
 must possess some hidden beauty, to  be unveiled
 when an exact solution is found.

\bigskip

\smallskip

\noindent
{\bf Acknowledgments.}

\smallskip

It is a pleasure to thank Ashoke Sen, Wati Taylor and Barton Zwiebach
for their constant interest  in our work
and for many detailed discussions  at various stages of this project.
This material is based upon work supported by the National Science
Foundation Grant No. PHY-9802484. Any opinions, findings,
and conclusions or recommendations expressed in this material
are those of the authors and do not necessarily reflect the views
of the National Science Foundation.

\appendix

\sectiono{The 
Numerical Algorithms}

In this appendix we explain some technical details
about the algorithms that we have used to
compute the universal star
products and find the tachyon solution 
in level truncation.

\subsection{Star Products from Conservation Laws}

The strategy for evaluating star products  using conservation laws
is explained in detail in \cite{0006240}.
Each Fock state in $\HH_{univ}^{(1)}$ can be represented as a string
of negatively moded ghost or Virasoro generators
acting on the zero-momentum tachyon $c_1 | 0 \ra$. The triple
product of three such states is evaluated
recursively by converting a negatively moded
generator  on one state space to a sum of positively
moded generators acting on all three state spaces, 
\ben \label{conslaw}
&& \la A_{-k}  \Phi_1,  \Psi_2,  \Psi_3 \ra   =  \\
&&  r_k   \la \Phi_1, \Psi_2, \Psi_3 \ra +
 \la   \sum_{n \geq 0}  \alpha_n^k  A_n \Phi_1 , \Psi_2, \Psi_3 \ra +
 \la   \Phi_1, \sum_{n \geq 0}  \beta_n^k  A_n \Psi_2,  \Psi_3 \ra +
 \la   \Phi_1,  \Psi_2, \sum_{n \geq 0}  \gamma_n^k  A_n  \Psi_3 \ra  \, ,\nonumber
\een
where $A_{-k}$ is any constructor symbol and the coefficients
appearing in this `conservation law' are computed from the geometry
of the Witten vertex \cite{0006240}. All triple products in $\HH^{(1)}_{univ} $
are thus reduced to the coupling $ \la c_1, c_1, c_1 \ra$
of three tachyons.  
\smallskip

Once the triple products are known,
star products are easily obtained by inverting the non-degenerate
bpz inner product. If $\{ \Psi_i \}$ is a Fock basis
for $\HH^{(1)}_{univ} $,  we define the dual basis
$\{  \Psi^i  \}$  of $\HH^{(2)}_{univ}$
by the bpz pairing
\be
\la  \Psi_i,  \Psi^j \ra  = \delta_i^j \,.
\ee
Then 
\be
\Psi_i * \Psi_j  \equiv     \sum_k  \la \Psi_i,  \Psi_j,  \Psi_k \ra \,  \Psi^k  \,.
\ee

We automated this algorithm on a C++ computer code. We 
briefly highlight some   features of our implementation:

\begin{itemize}
\item  We use the factorization of the star
product into matter and ghost sectors.  The algorithm
to find the triple products is executed separately in the two subsectors.

\item  We use cyclic and twist symmetry of the vertex
to reduce the computation to 
triple products  $\la \Psi_i , \Psi_j , \Psi_k \ra$ with a canonical
ordering $i \leq j \leq k$.

\item While in the matter sector the algorithm can be implemented
in a straightforward way, in the ghost sector we need
to face a slight complication. We are ultimately interested in triple
products of ghost number one states, but
the use of fermionic ghost conservation laws
necessarily brings us outside the ghost number one subspace. 
We found it  most efficient to use only conservation laws for the $b_{-k}$
oscillators. A single application of a $b$-ghost conservation law reduces
the evaluation of a $\la 1, 1, 1 \ra$ product (ghost number one in all three slots)
to a sum of terms of the form $\la 1, 1, 1 \ra$ 
{\it and}  $\la 0, 1, 2 \ra$.
Products of this latter type can be treated by applying 
a $b$-conservation law to the first state (of ghost number zero),
obtaining a sum of terms $\la 1, 1, 1\ra$ and again (after cyclic
rearrangement) $\la 0, 1, 2 \ra$.  It is easy to show that this algorithm always
terminates on the product of three tachyons.  So we see
that we only need to consider triple products of the form
$\la 0, 1, 2 \ra$  besides the standard products $\la 1, 1, 1 \ra$.

\item After each application of a conservation law,
the resulting triple products on the r.h.s of \refb{conslaw}
are processed using the Virasoro algebra or
the ghost commutation relations, till all states are reduced to
the Fock basis \refb{universal} with the canonical ordering
 $j_1 \geq \dots \geq j_p$, $k_1 \geq \dots \geq k_q$, $l_1 \geq \dots \geq l_r$. 
The evaluation of expressions like
$L_k  L_{-n_1} \cdots L_{-n_i}  c_1 | 0 \ra$, with
$n_1 \geq n_2 \cdots \geq n_i$ (and similarly for the ghosts)
 is thus a basic elementary operation. There is a
critical gain in efficiency in evaluating beforehand
all such expressions (up to the desired maximum level)
and reading the results from a file, rather then re-computing
them each time. The size of such a file grows
only linearly with the number of modes, whereas
the table of triple products grows cubically, so
this strategy is not problematic from the point
of view of memory occupation.

\end{itemize}

This algorithm can be easily extended to
evaluate more general star products 
of string fields belonging to a larger space
than $\HH_{univ}$, for example
the space relevant for tachyon lump solutions \cite{0005036}
or Wilson line solutions \cite{0007153}. One needs
to enlarge the algebra of matter operators and consider
the appropriate conservation laws.

\subsection{Solving the Equations of Motion}

Once all triple products at level $L$ have been computed,   
the evaluation of the star product of two Siegel gauge string fields
at level $L$ involves $O(N_L^3)$  algebraic operations.
It is clearly desirable to have an algorithm 
to solve the classical eom's that requires as few star
products as possible. We tried various options,  which can all be
represented as  a recursive procedure $\Psi^{(n+1)}=F(\Psi^{(n)})$, where $\Psi=F(\Psi)$
implies that $\Psi$ is a solution. 
\smallskip

The most obvious idea is to invert the kinetic term in Siegel gauge
and  define
\be
F(\Psi )=-\frac{b_{0}}{L_{0}} \, (\Psi *\Psi) \,.
\ee
Clearly this iteration cannot converge since $F(\lambda \Psi)=\lambda^{2}F(\Psi)$.
There is a simple way to fix  this problem,  defining
\be
\widetilde F(\Psi) = \left(  \frac{T[\Psi]}{   T [ F(\Psi)] } \right)^2
\,  F(\Psi) \, ,
 \ee
where $T[\Phi]$ indicates the coefficient of $c_1 |0 \ra$ in the
string field $\Phi$. Unfortunately the algorithm based on the recursion
$\widetilde F$  still 
fails to converge, and generically falls into stable two-cycles.
An improved recursion is
\be \label{alpharec}
F_\alpha(\Psi) = \alpha\, \Psi + (\alpha -1) \frac{b_{0}}{L_{0}} \Psi *\Psi \,
\ee
where $\alpha$ is a real number
which is chosen randomly
in some reasonable interval (say  $0.2 < \alpha < 0.8$)
 at each iteration step. This randomization 
breaks the cycles and  the algorithm converges to a unique solution  
in about 20-30 steps. (The algorithm stops when the eom's are satisfied with the same accuracy as the accuracy of  double-precision variables in C++, which have
15 significant digits). This algorithm is very robust with respect to the choice
of the starting point $\Psi_0$, in fact  at any given level $L$ we found 
only one non-trivial  solution.

\smallskip

A more efficient approach is the standard Newton
algorithm.  Recall that given a  system of $N$ algebraic equations in $N$ variables,
$f_i [ x_\alpha] =0$,  $1 \leq i, \alpha \leq N$,
the Newton recursion is
\be
x_{\alpha}^{(n+1)} =  x_{\alpha}^{(n)} - M^{-1}_{\alpha i}[x^{(n)}]  f_i [x^{(n)}]
\ee
where the matrix $M_{i \alpha} [x]$ is defined as
\be \label{M}
M_{i  \alpha} [ x]  \equiv \frac{\partial f_i }{ \partial x_{\alpha}} \,.
\ee
In our case, the truncated Siegel equations of motion are a system
of $N_L$ algebraic equations in $N_L$ variables (Table \ref{partitions})
and this method can be directly applied.  It is interesting
to write the Newton algorithm as a recursion
for the Siegel string field itself. One finds the compact
expression
\be
\Psi^{(n+1)} = Q_{\Psi^{(n)}} ^{-1}(\Psi^{(n)} * \Psi^{(n)} ) \,.
\ee
Here the operator $Q_\Psi$ is defined by
\be
Q _\Psi \, \Phi \equiv Q_B \Psi + \Psi * \Phi  + \Phi* \Psi
\ee
for any ghost number one string field $\Phi$. The inverse operator $Q^{-1}_ \Psi$ is 
naturally defined by projecting onto the Siegel subspace. In other terms,
 for any ghost number two string field
 $\Sigma$, we look for the ghost number one string field $Q^{-1}_\Psi   \Sigma$ 
that obeys
\be
b_0 \, (  Q^{-1} _\Psi  \Sigma)  = 0 \, , \quad              c_0 b_0 \, \Sigma =  c_0 b_0\,     Q_\Psi (Q^{-1}_\Psi) \, \Sigma \,.
\ee
The operator $Q_\Psi$ has a natural physical interpretation:
 If $\Psi$ is a solution of the OSFT eom's, then
$Q_\Psi$ is  the new BRST operator obtained expanding the OSFT
action around $\Psi$.  Thus as we approach
the fixed point of the Newton recursion,
$Q_{\Psi^{(n)} }$ becomes a better and better
approximation to the BRST operator around the tachyon vacuum.

\smallskip

In level-truncation,  the action of the operator $Q_\Psi$ in
the Siegel subspace is represented by an $N_L \times N_L$ matrix.
Since there is an order $O(N_L^3)$ algorithm to invert
a matrix, the Newton recursion is not significantly more 
time-expensive than the evaluation of a single star product.
The Newton algorithm is very fast, effectively squaring the accuracy at each step,
and the solution is reached in four or five iterations. 
On our pc, the complete algorithm (computing the vertices from scratch
and finding the tachyon solution) takes less than 10 seconds at level (10,20),
and less than a minute at level (12,36)!
There is however
a rather critical dependence on the initial conditions: one finds convergence
only from a starting point sufficiently close to the solution
(it is enough to take {\it e.g.} $\Psi^{(0)} = 0.5\, c_1 |0 \ra$).

\smallskip

 We compared the solution
obtained with the Newton method with the solution found with the alternative algorithm described above, finding exact agreement up to level 16.  (At level 18 the recursion
\refb{alpharec} runs too slowly  on our pc).     This gives a 
strong check on the correctness of the solutions. As another check,
we compared our results at level (10,20) with the results
of  Moeller and Taylor \cite{MT}\footnote{We thank the authors of \cite{MT}
for making their full results available to us.}, finding agreement to the tenth
significant digit.

\subsection{Tachyon Effective Action}

To compute the tachyon effective action, we write the
string field as
\be \label{psitilde}
\Psi_L = T \, c_1 | 0 \ra + \widetilde \Psi_L \, ,
\ee
where $\widetilde \Psi_L$ contains all the modes
up to level $L$, except $c_1 | 0 \ra$.
For a given numerical value of of the variable $T$,
we solve the classical OSFT equations of motion for all the higher modes,
using the Newton method.  This gives  
$\Psi_L[T] = T c_1 | 0 \ra + \widetilde \Psi_L [ T]$
as a function of $T$.  Plugging $\Psi_L[T]$ into the OSFT
action\footnote{
More precisely, $V_L(T) \equiv f ( \Psi_L[T])$,
where $f(\Psi)$ is defined in \refb{fPsi}.}
we obtain the effective tachyon potential $V_L(T)$.

\smallskip

The Newton algorithm that finds the solution $\widetilde \Psi_L [T]$ fails
to converge if the variable $T$ is outside an interval
$[T_L^{min}, T_L^{max}]$. We find
for example $T^{min}_{16} \sim -0.1 $ and $T^{max}_{16} \sim 0.7 $
(notice that both the tachyon vacuum and the
perturbative vacuum are safely inside the convergence region).
The failure of the numerical algorithm can be explicitly traced to the existence
of other branches in the tachyon effective action. This phenomenon has been 
studied in \cite{MT, ET2}, 
where it has also been related to the non-perturbative failure
of the Siegel gauge condition.  In this paper we only need $V_L(T)$
in an interval around the non-perturbative vacuum, which we take
to be $0.54 \leq T \leq 0.55$. We postpone 
a more detailed investigation of the global behavior of the tachyon potential.

\section{Some Further Numerical Data}

\vspace{3cm}

\label{further}

\begin{table}[h]
\begin{tabular}{|l|l|l|l|l|}
\hline
&$L=4$&$L=6$&$L=8$&$L=10$ \\
\hline
$M=4$&0.54839904&0.54849677&0.54814406&0.54777626 \\
\hline
$M=6$&           &0.54793242&0.54711284&0.54639593\\
\hline
$M=8$&           &           &0.54705245&0.54626520\\
\hline
$M=10$&          &           &           &0.54626093\\
\hline
\hline
&$L=12$&$L=14$&$L=16$&$L=18$ \\
\hline
$M=4$&0.54745869&0.54719393&0.54697362&0.54678893 \\
\hline
$M=6$&0.54581507&0.54534684&0.54496539&0.54465027 \\
\hline
$M=8$&0.54561932&0.54509452&0.54466463&0.54430807 \\
\hline
$M=10$&0.54560864&0.54507682&0.54464004&0.54427703 \\
\hline
$M=12$&0.54560809&0.54507524&0.54463714&0.54427267 \\
\hline
$M=14$&          &0.54507515&0.54463683&0.54427204 \\
\hline
$M=16$&           &          &0.54463682&0.54427198 \\
\hline
$M=18$&           &          &          &0.54427196 \\
\hline
\end{tabular}
\caption{Estimates $T^{(M)}_L$ for the tachyon vev
obtained from extrapolations of the effective tachyon  potential,
at various orders $M$ and for $L \leq 18$.
Data in the $(L,3L)$ scheme. By
definition, the diagonal entries $T^{(M=L)}_L$ coincide with the exact
computation from direct level-truncation at level $(L, 3L)$
(Table \ref{sample}).\label{Mtachyon3L}  }
\end{table}

\begin{table}
\begin{tabular}{|r|r|r|r|}
\hline
Matter&Ghost&$L=\infty$&$L=18$ \\
\hline
      &              &.5405&0.5443 \\
\hline
      &$b_{-1}c_{-1}$&-0.2248&-0.2205 \\
\hline
$L^m_{-2}$&            &0.05721&0.05726 \\
\hline
      &$b_{-1}c_{-3}$&0.05928&0.05879 \\
\hline
      &$b_{-2}c_{-2}$&0.03650&0.03564 \\
\hline
      &$b_{-3}c_{-1}$&0.01976&0.01960 \\
\hline
$L^m_{-2}$&$b_{-1}c_{-3}$&0.008627&0.008197 \\
\hline
$L^m_{-4}$&           &-0.005049&-0.005079 \\
\hline
$L^m_{-2}L^m_{-2}$&          &-0.000681&-0.000661 \\
\hline
      &$b_{-1}c_{-5}$&-0.03091&-0.03076 \\
\hline
      &$b_{-2}c_{-4}$&-0.01976&-0.01941 \\
\hline
      &$b_{-3}c_{-3}$&-0.01152&-0.01151 \\
\hline
      &$b_{-2}b_{-1}c_{-2}c_{-1}$&-0.008626&-0.008316 \\
\hline
      &$b_{-4}c_{-2}$&-0.00988 &-0.009704 \\
\hline
      &$b_{-5}c_{-1}$&-0.00618 &-0.006152 \\
\hline
$L^m_{-2}$&$b_{-1}c_{-3}$&-0.003702&-0.003605 \\
\hline
$L^m_{-2}$&$b_{-2}c_{-2}$&-0.003186&-0.003056 \\
\hline
$L^m_{-2}$&$b_{-1}c_{-1}$&-0.001234&-0.001202 \\
\hline
$L^m_{-3}$&$b_{-1}c_{-2}$&-0.000076&-0.0000775 \\
\hline
$L^m_{-3}$&$b_{-2}c_{-1}$&-0.000038&-0.0000387 \\
\hline
$L^m_{-4}$&$b_{-1}c_{-1}$&-0.0012&-0.001242 \\
\hline
$L^m_{-2}L^m_{-2}$&$b_{-1}c_{-2}$&-0.000248 &-0.000215 \\
\hline
$L^m_{-6}$&        &0.001434&0.001446 \\
\hline
$L^m_{-3}L^m_{-3}$&       &0.0000075&0.0000075 \\
\hline
$L^m_{-4}L^m_{-2}$&       &0.000311&0.000310 \\
\hline
$L^m_{-2}L^m_{-2}L^m_{-2}$&        &-0.0000049&-0.0000065 \\ 
\hline
\end{tabular}
\caption{ \label{asymptotic} Asymptotic values 
for the first coefficients of the tachyon condensate string field,
compared with the $L=18$ data.  The $L =\infty$ results
are obtained from the $M=16$ extrapolation procedure
based on the effective tachyon potential. 
Data in the $(L,3L)$ scheme.}
\end{table}

\begin{table}
\begin{tabular}{|l|l|l|l|l|}
\hline
&$L=4$&$L=6$&$L=8$&$L=10$ \\
\hline
$M=4$&-0.20567285&-0.21119493&-0.21392087&-0.21552208 \\
\hline
$M=6$&           &-0.21181486&-0.21499106&-0.21690691\\
\hline
$M=8$&           &           &-0.21502535&-0.21697620\\
\hline
$M=10$&          &           &           &-0.21698254\\
\hline
\hline
&$L=12$&$L=14$&$L=16$&$L=18$ \\
\hline
$M=4$&-0.21656852&-0.21730328&-0.21784642&-0.21826369 \\
\hline
$M=6$&-0.21818110&-0.21908711&-0.21976325&-0.22028663 \\
\hline
$M=8$&-0.21827982&-0.21920964&-0.21990503&-0.22044413 \\
\hline
$M=10$&-0.21829559&-0.21923573&-0.21994128&-0.22048993 \\
\hline
$M=12$&-0.21829570&-0.21923600&-0.21994171&-0.22049051 \\
\hline
$M=14$&          &-0.21923603&-0.21994180&-0.22049069 \\
\hline
$M=16$&           &          &-0.21994181&-0.22049069 \\
\hline
$M=18$&           &          &          &-0.22049069\\
\hline
\end{tabular}
\caption{ Estimates for the vev of $c_{-1} | 0 \ra$, 
obtained from extrapolations of the effective tachyon  potential,
at various orders $M$ and for $L \leq 18$.
Data in the $(L,3L)$ scheme. 
By definition, the diagonal entries  coincide with the exact
computation from direct level-truncation at level $(L,3L)$ (Table \ref{sample}).
\label{Mu3L}    }
\end{table}

\begin{table}
\begin{tabular}{|l|l|l|l|l|}
\hline
&$L=4$&$L=6$&$L=8$&$L=10$ \\
\hline
$M=4$&0.056923526&0.057062755&0.057068423&0.057045668 \\
\hline
$M=6$&           &0.057143493&0.057209039&0.057229308 \\
\hline
$M=8$&           &           &0.057214101&0.057239805\\
\hline
$M=10$&          &           &           &0.057241066\\
\hline
\hline
&$L=12$&$L=14$&$L=16$&$L=18$ \\
\hline
$M=4$&0.057018045&0.056991677&0.056968053&0.056947298 \\
\hline
$M=6$&0.057233609&0.057231744&0.057227479&0.057222387 \\
\hline
$M=8$&0.057248895&0.057251070&0.057250189&0.057247946 \\
\hline
$M=10$&0.057252093&0.057256442&0.057257742&0.057257584 \\
\hline
$M=12$&0.057252005&0.057256182&0.057257253&0.057256834 \\
\hline
$M=14$&          &0.057256190&0.057257279&0.057256887 \\
\hline
$M=16$&           &          &0.057257279&0.057256886 \\
\hline
$M=18$&           &          &          &0.057256885\\
\hline
\end{tabular}
\caption{ Estimates for the vev of $L^m_{-2} c_{1} | 0 \ra$, 
obtained from extrapolations of the effective tachyon  potential,
at various `orders'  $M$ and for $L \leq 18$.
Data in the $(L,3L)$ scheme. By
definition, the diagonal entries  coincide with the exact
computation from direct level-truncation at level $(L, 3L)$ (Table \ref{sample}).
 \label{Mv3L}   } 
\end{table}

\begin{table}
\begin{tabular}{|r|r|r|r|r|r|r|}
\hline
& $L=10$ & $ L =12$  & $L=14$ & $L=16$ & $L=18$  &   $conj$\\  
\hline
$r_{9,1}$ & 0.0259407 & 0.0261085 & 0.0262255 & 0.0263037 & 0.0263594 & 0.027063 \\
\hline   
$r_{7,3} $    &     0.0104886 & 0.0104825 & 0.0104948&0.0105063&  0.0105159
&  0.0105868 \\
\hline
$r_{5,5}$&  0.00658638 &  0.0065773 & 0.00658224 &0.00658778& 0.00659265 & 0.0066192  \\
\hline
$r_{11, 1}$  & &  -0.0200117 &  -0.0201181 &  -0.0201948 &  -0.0202469 & -0.0208326 \\
\hline
$r_ {9, 3}$  & & --0.00818159 &  -0.00817378 &  -0.00818 &  -0.00818657 &  -0.00824041\\
\hline
 $r_{7, 5}$ & & -0.00525767 &  -0.00524794 &  -0.00524929 &  -0.00525185    & -0.00526601 \\
\hline
$r_{13, 1}$ & & & 0.0161045 & 0.0161778 &  0.0162318&  0.0167396  \\
\hline
$r_{11, 3}$ & & & 0.00662999 &  0.00662276 & 0.0066262  & 0.0066689 \\
\hline
$r_{9, 5}$  & & &  0.00431978 &  0.00431134 &  0.00431141 &  0.00431915 \\
\hline
$r_{7, 7}$   & & & 0.00315946 & 0.00315272 &  0.00315241&  0.0031549 \\
\hline 
$r_{15, 1}$& & &  & -0.0133607 &  -0.0134141 &  -0.013871\\
\hline
$r_{13, 3}$ & & & &  -0.0055259 &  -0.00551968   & -0.005553 \\
\hline
$r_{11, 5}$& & & &  -0.00363327 &  -0.00362626   & -0.003629 \\
\hline
\end{tabular}
\caption{\label{higherR} $(L,3L)$ numerical results for the pattern coefficients
$r_{n,m}$ for the tachyon condensate.     }
\end{table}

\end{document}